\documentclass[pra,twocolumn,groupedaddress,showpacs]{revtex4-1}

\usepackage{amsmath,amssymb}
\usepackage{amsthm}
\usepackage{graphicx}
\usepackage{epstopdf}

\newtheorem{definition}{Definition}
\newtheorem{proposition}[definition]{Proposition}
\newtheorem{lemma}[definition]{Lemma}

\newtheorem{theorem}[definition]{Theorem}
\newtheorem{corollary}[definition]{Corollary}
\newtheorem{conjecture}[definition]{Conjecture}
\newtheorem{postulate}[definition]{Postulate}

\newtheorem{remark}[definition]{Remark}
\newtheorem{example}[definition]{Example}
\newtheorem{protocol}[definition]{Protocol}
\newtheorem{question}[definition]{Question}

\def\cites#1{XX Need reference XX}
\long\def\ca#1\cb{}

\def\dyad#1#2{|#1\rangle\langle#2|}

\def\ket#1{|#1\rangle }

\def\ra{\rightarrow }
\def\cC{{\cal C}}

\def\EC{{\cal E}}

\def\cH{{\cal H}}

\def\ox{\otimes}

\def\bcj{\begin{conjecture}}
\def\ecj{\end{conjecture}}
\def\bcr{\begin{corollary}}
\def\ecr{\end{corollary}}
\def\bd{\begin{definition}}
\def\ed{\end{definition}}
\def\bea{\begin{eqnarray}}
\def\eea{\end{eqnarray}}
\def\bem{\begin{enumerate}}
\def\eem{\end{enumerate}}
\def\bex{\begin{example}}
\def\eex{\end{example}}
\def\bptl{\begin{protocol}}
\def\eptl{\end{protocol}}
\def\bim{\begin{itemize}}
\def\eim{\end{itemize}}
\def\bl{\begin{lemma}}
\def\el{\end{lemma}}
\def\bpf{\begin{proof}}
\def\epf{\end{proof}}
\def\bpp{\begin{proposition}}
\def\epp{\end{proposition}}
\def\bqu{\begin{question}}
\def\equ{\end{question}}
\def\br{\begin{remark}}
\def\er{\end{remark}}
\def\bt{\begin{theorem}}
\def\et{\end{theorem}}
\def\btb{\begin{tabular}}
\def\etb{\end{tabular}}

\def\diag{\mathop{\rm diag}}

\def\max{\mathop{\rm max}}

\long\def\ca#1\cb{}

\begin{document}


\title{Implementation of bipartite or remote unitary gates with repeater nodes}

\author{Li Yu$^1$}
\email{yupapers@sina.com}

\author{Kae Nemoto$^1$}
\affiliation{$^1$National Institute of Informatics, 2-1-2 Hitotsubashi, Chiyoda-ku, Tokyo 101-8430, Japan}

\begin{abstract}
We propose some protocols to implement various classes of bipartite unitary operations on two remote parties with the help of repeater nodes in-between. We also present a protocol to implement a single-qubit unitary with parameters determined by a remote party with the help of up to three repeater nodes. It is assumed that the neighboring nodes are connected by noisy photonic channels, and the local gates can be performed quite accurately, while the decoherence of memories is significant. A unitary is often a part of a larger computation or communication task in a quantum network, and to reduce the amount of decoherence in other systems of the network, we focus on the goal of saving the total time for implementing a unitary including the time for entanglement preparation. We review some previously studied protocols that implement bipartite unitaries using local operations and classical communication and prior shared entanglement, and apply them to the situation with repeater nodes without prior entanglement. We find that the protocols using piecewise entanglement between neighboring nodes often require less total time compared to preparing entanglement between the two end nodes first and then performing the previously known protocols. For a generic bipartite unitary, as the number of repeater nodes increases, the total time could approach the time cost for direct signal transfer from one end node to the other. We also prove some lower bounds of the total time when there are a small number of repeater nodes. The application to position-based cryptography is discussed.
\end{abstract}

\date{Version of \today}
\pacs{03.67.Hk, 03.67.Dd, 03.67.Ac, 03.67.Bg}

\maketitle



\section{Introduction}
\label{sct1}

For quantum computation to outperform classical computers, it is necessary for the quantum computer to have a large scale. But, local quantum computers may be limited in size, thus distributed quantum computation may be needed. The input data and the output states may be needed at different locations, dependent on the actual needs.
Various nonlocal communication or cryptographic tasks, such as state merging, quantum fingerprinting, quantum secret sharing, quantum voting, etc., may also involve quantum computation on each party with sending of quantum information among the parties, or manipulating a shared quantum state. Hence, the ability to perform a nonlocal gate on two remote parties is beneficial for both quantum computation and quantum cryptology.

Long-distance transmission of quantum state, such as that using photons, is often subject to severe errors such as photon loss. Since quantum repeaters may help overcome such difficulty, they have been the subject of theoretical and experimental studies \cite{dlcz01,msd12,Bernien13,ywbp16} (also see reviews \cite{gt07,ssr11,mat15}), usually with the intended application of establishing entanglement or shared classical keys between remote parties. But the use of repeaters in the more general task of doing an arbitrary quantum operation has not been studied thoroughly. The nonlocal unitaries are a simplest class of nonlocal quantum operations, since the nonlocal quantum operations, modeled as completely positive trace-preserving maps on the two or more parties, can be implemented using nonlocal unitaries followed by local measurements. (See Sec.~\ref{ssct:3.5} for details.) In this paper, we study the use of quantum repeaters to perform nonlocal unitary gates, and a class of local single-qubit unitary gates with its parameter determined by a remote party. We call these two classes of gates as \emph{bipartite unitaries} and \emph{remote unitaries}, respectively. Our protocols are based on some previously studied protocols that implement bipartite unitaries using local operations and classical communication (LOCC) and prior shared entanglement between two parties, as well as a protocol mentioned in \cite{hpv02} for implementing remote single-qubit unitary gates with parameters controlled by another party. Some general thoughts about how the repeaters are used in the bipartite unitary protocols is in Sec.~\ref{sec:discussion}.

The total time cost for implementing a unitary is important because shorter time means less decoherence for the systems in the network which may or may not be directly acted on by the unitary. The time for entanglement preparation is included for two reasons: First, memories have decoherence so they might not store entangled states very well, thus fresh entanglement may need to be generated before doing the unitary. Second, the entangled states needed by our protocols are among many different systems at different locations, and the amount of entanglement needed might not be known beforehand, and thus it may be costly to maintain entanglement between all pairs of nodes in a network at all times. As discussed at the beginning of Sec.~\ref{ssct:3.2}, the time cost of a protocol for implementing a unitary in this paper refers to the time in the idealized scenario of infinite copies of local ancillary systems for establishing entanglement between nodes (abbreviated as \emph{local setups}). But, since in practice we have finite copies of the local setups, the overall operation implemented after one run of the protocol is in general a quantum operation (i.e., a completely-positive trace-preserving map) that approximates the ideal quantum operation corresponding to the target unitary. The error in the approximate implementation would be smaller and smaller if we increase the number of local setups. In the limit of an infinite number of local setups for generating entanglement, the unitary would (ideally) be implemented exactly.

We show that the total time cost for implementing a bipartite unitary on two end nodes (denoted as $A$ and $B$ throughout the paper) of a repeater array is strictly greater than $\frac{L}{c}$ under some assumptions detailed in Sec.~\ref{sct2}, where $L$ is the distance between the two end nodes, and $c$ is the speed of light in the relevant medium. The total time cost of our protocols approaches $\frac{L}{c}$ from above as the number of repeater nodes increases.

The fact that the total time cost is strictly greater than $\frac{L}{c}$ has some implications for position-based quantum cryptography \cite{LL11,Malaney10,kms11,Buhrman14,bk11,unruh14,cl15}, which is the quantum version of classical position-based cryptography \cite{Chandran09}, and is related to the topic of instantaneous nonlocal measurement \cite{Vaidman03,Clark10}. In position-based cryptography, we are usually interested in the task of position verification (or tagging), which is for a prover to prove to some remote verifiers that he or she is at a particular spatial location. Quantum position verification with two verifiers may become secure when the assumptions listed in Sec.~\ref{ssct:3.4} and the conditions in Postulate~\ref{postu:position_verification} both hold.

The structure of the paper is as follows. In Sec.~\ref{sct2} we mention a few previously studied protocols for implementing bipartite nonlocal unitaries, and some general considerations about the types of protocols to be studied in this paper. In Sec.~\ref{sct3} we present the main results, including the application to position-based quantum cryptography. In Section~\ref{sec:discussion}, we review the general way the repeater nodes are used in the bipartite unitary protocols, and compare it to how repeaters are used in generating long-range entanglement in the literature. The conclusions and some open problems are given in Sec.~\ref{sct4}.

\section{Preliminaries}
\label{sct2}

{\bf Notations.}
Suppose we hope to implement a unitary $U: \cH_{AB}\rightarrow\cH_{AB}$ on the systems $A$ and $B$.  The nodes of the network where $A$ and $B$ are located are called the end nodes. There are $n$ other nodes in the network which are called repeater nodes. The distance between the two end nodes is denoted as $L$. Let $I_A$ ($I_B$) denote the identity operator on the Hilbert space $\cH_A$ ($\cH_B$). Let $d_A$ and $d_B$ be the dimensions of $\cH_A$ and $\cH_B$, respectively.

We define some terms about relations between time periods of the steps in the protocols. Since our protocols allow doing multiple things at different locations simultaneously, a step of the protocol may involve several logical steps acting on different systems at the same time. Each such logical step can be called a process, for the purpose of the definitions below. ``Process 1 coincides with process 2'' means that the two processes start simultaneously and last for the same length of time. If the time interval for process 1 is within that for process 2, we may say that (the time interval for) process 1 is contained in the time interval for process 2. And ``process 1 partially overlaps with process 2'' means no one time interval need to be completely within the other.

We use ebits and c-bits to measure the entanglement cost and classical communication cost of a protocol, respectively. The entanglement contained in a maximally entangled pure state of Schmidt rank $N$ is regarded as $\log_2 N$ ebits. If the classical message is a signal among $N$ equally possible signals, the amount of classical communication in this message is regarded as $\log_2 N$ c-bits.

In this section we will introduce a few known protocols that implement bipartite unitaries or a remotely determined local unitary using entanglement and LOCC. Since some of the protocols work only for special types of unitaries, we shall first introduce the types of unitaries before giving the description of the protocols. The variants to these protocols under the presence of the repeater nodes will be introduced in Sec.~\ref{sct3} below.

{\bf Assumptions.} In this paper we are mainly interested in the total time cost of implementing a bipartite unitary or a remotely determined local unitary. To make a fair comparison between the total time cost of different protocols, some general assumptions are as follows:\\
(a) The quantum channels between nodes are noisy, but the local operations (including local quantum gates and local measurements) are regarded as having no errors and taking no time to implement, and classical communication is error-free.\\
(b) The speed of light is uniform in the relevant medium and is denoted as $c$.\\
(c) All nodes do not have prior knowledge about the time of the actual run of the protocol, and are notified of the start of the protocol at the same time.\\
(d) The quantum memories of all nodes have finite decoherence time.\\
(e) The number of local gates and measurements should not be much larger than in the original protocols without using repeaters.

The assumption (a) implies the following: Since the information about the input of the unitary should better not be lost, the systems containing information about the input should not be directly transmitted in the channels, but should rather be transmitted by teleportation or similar schemes with the help of entanglement prepared in previous steps of the protocol. (The method of teleportation has the advantage that the entanglement can be re-prepared in case preparation fails, without affecting the input quantum state.)

The assumptions (c) and (d) together imply that the nodes cannot have shared entanglement before the protocol. Hence, the time cost of the protocol needs to include the time for entanglement preparation (between the relevant pairs of nodes).

All the protocols in this paper satisfy assumption (e), but we still list this assumption to exclude possible schemes with unacceptably large entanglement consumption, such as a few fast unitary protocols (or called ``instantaneous nonlocal quantum computation'') in the literature \cite{Buhrman14,bk11} applied to generic bipartite unitaries. In general, such fast protocols implement the target unitary approximately using a very large amount of entanglement. We exclude the use of such fast protocols for generic unitaries, but still use some of the known fast protocols for specific classes of unitaries (see Protocols 5 and 6 below).

{\bf Type of (expansions of) unitaries.}

Some of the protocols discussed in this paper are for implementing any bipartite unitary $U$, but some others are for implementing special types. The types are as follows (see \cite{ygc10}), but note that the types (b) and (c) are merely types of expansions of unitaries; these two ``types'' actually both contain all bipartite unitaries.

(a) Controlled unitaries.
\bea\label{u:control}
U=\sum_{j=1}^N P_j \ox V_j,
\eea
where $\{P_j\}$ is a set of mutually orthogonal projectors of integer rank, and $\sum_j P_j=I_A$, and $V_j$ are arbitrary unitary operators on $\cH_B$. The integer $N$ is called the number of terms in the controlled unitary $U$.

(b) Double-group unitaries.
\bea\label{u:double_group}
U=\sum_{f\in G} c(f) V_A(f)\ox T_B(f),
\eea
where $V_A(f)\ox T_B(f)$ form a projective representation of a finite group $G$, and $c(f)$ are complex numbers. Hence each of the two sets $\{V_A(f)\}$ and $\{T_B(f)\}$ is a projective representation of the group $G$. The double-group form can be viewed as a generic form of bipartite unitaries, because any bipartite unitary on $d_A\times d_B$ system can be expanded using the form \eqref{u:double_group} with a group of size $d_A^2 d_B^2$, with the group elements represented by $X_A^j Z_A^k \ox X_B^l Z_B^m$, where the generalized Pauli operators $X$ and $Z$ are defined by
\bea\label{eq:xzdef}
X&=&\sum_{k=0}^{N-1}\dyad{(k-1)\mbox{ mod }N}{k},\notag\\
Z&=&\sum_{k=0}^{N-1} e^{2\pi i k/N}\dyad{k}{k}
\eea
for an $N$-dimensional Hilbert space, and the $j,k,l,m$ are integer labels.

(c) Single-group unitaries. The expansion of $U$ involves a group representation only on one side, i.e.
\bea\label{u:single_group}
U=\sum_{f\in G} V_A(f)\ox W_B(f),
\eea
where $V_A(f)$ form a projective representation of a finite group $G$, and $W_B(f)$ are arbitrary operators on $\cH_B$ satisfying that the $U$ given above is unitary. Any bipartite unitary on $d_A\times d_B$ system can be expanded in the form \eqref{u:single_group}, with a group of size $d_A^2$, where the group elements are represented by $X_A^j Z_A^k$, and the $X$ and $Z$ are defined in \eqref{eq:xzdef}. For generic unitaries, this expansion is more efficient than the form \eqref{u:double_group} in that the number of terms is smaller, which means the entanglement cost in a protocol for implementing the unitary would be smaller, but for our protocols in this paper with at least one repeater node, the time cost would be higher because the form \eqref{u:single_group} is less structured than the form \eqref{u:double_group} in that there are no strong requirements on $W_B(f)$.

Now we introduce some types of protocols that do not use repeater nodes, and their variants which use repeater nodes will be discussed in the next section.

{\bf Protocol 1.} The two-way teleportation protocol. Suppose the system $A$ has smaller size among the two input systems. The protocol involves teleporting the system $A$ to the location of the other party, performing the unitary there, and teleporting the system $A$ back to the original location. The protocol uses entangled ancillae $a$ and $b$ which are at the locations of the input systems $A$ and $B$, respectively (this holds for the Protocols 1 through 7). The entangled resource needed is two copies of $d_A\times d_A$ maximally entangled states of the form $\frac{1}{\sqrt{d_A}}\sum_{j=1}^{d_A} \ket{jj}$. The classical communication cost of this protocol, measured in c-bits, is twice the number of ebits required by the same protocol. The same is true for the Protocols 2 through 7.

{\bf Protocol 2.}  The protocol for implementing controlled unitaries in Sec.~III of \cite{ygc10}. The form of $U$ is given above in Eq.~\eqref{u:control}. This will be called ``the basic controlled-unitary protocol'' throughout this paper. The protocol uses a maximally entangled state $\frac{1}{\sqrt{N}}\sum_{j=1}^N \ket{jj}$. So the entanglement cost is independent of $d_A$ and $d_B$ when $N$ is fixed. The figure for this protocol will appear as Fig.~\ref{fgr_cont} in Sec.~\ref{ssct:3.2}, which also discusses the preparation of the entangled state on $ab$.

{\bf Protocol 3.}  The double-group type protocol in Sec.~IV of \cite{ygc10}. The form of $U$ is given above in Eq.~\eqref{u:double_group}. The protocol uses a maximally entangled state $\frac{1}{\sqrt{d}}\sum_{j=1}^d \ket{jj}$, where $d=\vert G\vert$. So the entanglement cost is independent of $d_A$ and $d_B$ when $G$ is fixed. The figure for this protocol will appear as Fig.~\ref{fgr_group} in Sec.~\ref{ssct:3.2}. As mentioned previously, the type of unitary $U$ in \eqref{u:double_group} could include any bipartite unitary.

{\bf Protocol 4.} The general group-type protocol in \cite[Sec.~IV]{ygc10}. The form of $U$ is given above in Eq.~\eqref{u:single_group}. The protocol uses a maximally entangled state $\frac{1}{\sqrt{d}}\sum_{j=1}^d \ket{jj}$, where $d=\vert G\vert$. The figure for this protocol is given in \cite[Fig.~8]{ygc10}, and also appears as Fig.~\ref{fgr_gengroup} in Sec.~\ref{ssct:3.2}. As mentioned previously, the type of unitary $U$ in \eqref{u:single_group} could include any bipartite unitary.

{\bf Protocol 5.}  The fast double-group type protocol in \cite[Sec. III]{ygc12} with the circuit shown in Fig.~\ref{fgr_fastgroup} in Sec.~\ref{ssct:3.2}. The protocol uses only one round of parallel classical communication in two opposite directions. Any protocol that implements bipartite unitaries and has such property is called a fast unitary protocol. Generally, any fast unitary protocol is also called ``instantaneous quantum computation'' in the literature \cite{Buhrman14,bk11,cl15,Speelman15,Broadbent15}. For the current Protocol 5, we consider a special type of fast unitary protocol. The form of the unitary is formally the same as in Eq.~\eqref{u:double_group} but with some additional constraints on the coefficients $c(f)$:
\bea\label{u:double_group2}
U=\sum_{f\in G} c(f) V_A(f)\ox T_B(f),
\eea
where $V_A(f)\ox T_B(f)$ form a projective representation of a finite group $G$, and $c(f)$ are complex numbers subject to special conditions. The protocol uses a maximally entangled state $\frac{1}{\sqrt{d}}\sum_{j=1}^d \ket{jj}$, where $d=\vert G\vert$, but note that the $G$ may be a larger group than in Protocol 3 for the same unitary $U$.

We do not discuss the fast version of Protocol 4, because in the generic case it is covered by Protocol 5, see Proposition~\ref{prop1} below. But there are special cases in which a fast version is possible. Such cases that we know of all satisfy that the $W_B(f)$ are not all invertible.
An example is illustrated by Protocol 6 below (with the role of the two parties exchanged). Another type of example is the tensor product of a unitary in Protocol 6 with a unitary implementable by Protocol 5.

{\bf Protocol 6.} The fast controlled-Abelian-group protocol in \cite[Sec.~II]{ygc12} or more generally, the fast controlled-group protocol in \cite{Yu11} with the circuit shown in Fig.~\ref{fgr_ctgroup} in Sec.~\ref{ssct:3.2}. The $U$ is of the form \eqref{u:control} but with the $V_j$ forming a subset of a projective representation of a finite group $G$. The protocol uses an entangled resource state of the same form as in Protocol 5. In the case of fast controlled-Abelian-group protocol in \cite[Sec.~II]{ygc12}, the $U$ is equivalent to the form of \eqref{u:double_group2} under local unitaries, thus can be implemented using Protocol 5.

{\bf Protocol 7.} This is not a single protocol, but refers to the family of all possible fast protocols that implement $U$ exactly or approximately. The two parties $A$ and $B$ perform local operations including measurement, and send the measurement outcomes to each other simultaneously, and then perform some local operations in order to implement the bipartite unitary $U$. This class of protocols include the ``instantaneous quantum computation'' protocols discussed in the position-based cryptography literature, e.g. the protocols in \cite{Buhrman14,bk11,cl15,Speelman15}. It also includes the fast protocol for implementing two-qudit Clifford unitaries, which is generalized from the protocol shown in \cite[Fig. 2]{gc99} by the following changes: change the target gate from a controlled-NOT (CNOT) gate to any two-qudit Clifford gate denoted $U$, and replace the initial state $\ket{\chi}$ by $\frac{1}{d}\sum_{j=1}^d\sum_{k=1}^d \ket{j}_a U(\ket{j}_A \ket{k}_B)\ket{k}_b$ (the systems $a$,$A$,$B$,$b$ correspond to the four middle lines of \cite[Fig. 2]{gc99}, in the up-to-down order), and replace the local Bell measurements by generalized Bell measurements, and replace the Pauli gates by generalized Pauli gates. [The generalized Pauli group ${\cal P}_1$ on one qudit is generated by the two operators in \eqref{eq:xzdef}, and the generalized Pauli group on $n$ qudits, ${\cal P}_n$, is defined as the $n$-fold tensor product of ${\cal P}_1$; the generalized Clifford group on $n$ qudits is defined as the set of operators $C$ that satisfy $C {\cal P}_n C^\dag={\cal P}_n$.] The reason such protocol works is that the generalized Clifford operators map the generalized Pauli operators to the generalized Pauli operators. This protocol can be extended to the case that the $U$ is a bipartite Clifford operator on $m+n$ qudits, where the first $m$ qudits are in $A$ and the remaining $n$ qudits are in $B$. We shall later discuss the variant of Protocol 7 in the case of two repeater nodes (Protocol 7.2). Although the bipartite Clifford unitaries are a special class of bipartite unitaries, this set of gates has many useful properties, and when aided by all one-qubit unitaries (or even stronger, by a single generic one-qubit unitary \cite{Shi03}) they become universal for quantum computation.

{\bf Protocol 8.}
The protocol for implementing a local single-qubit unitary $U=\diag(e^{i\theta},e^{-i\theta})$ with parameter $\theta$ determined by a remote party. In particular, we consider the protocol for Bob to perform $U$ on system $B$ with the rotation angle $\theta$ determined by Alice at location $A$, as shown in \cite[Fig.~3]{hpv02}. As $\theta$ is a continuous parameter, we assume that Alice cannot directly send the information about $\theta$ to Bob via a classical channel. This can also be because Alice hopes to keep $\theta$ secret, which makes sense in our later analysis of Protocol 8.1 where we are only concerned with some particular values of $\theta$. As shown in \cite{hpv02}, there is a protocol that uses one ebit and two c-bits, where the two c-bits are sent in opposite directions. The protocol uses entangled ancillae $a$ and $b$ which are at the locations $A$ and $B$, respectively.

For the following approximate protocol, we do not discuss its variants, since adding nodes does not reduce the total time cost under this protocol.

{\bf Protocol 9.} Assume that the error rates in the channels are such that the quantum capacity of the channel from one node to a neighboring node is always nonzero. There is a repeater node $C$ at the middle point between $A$ and $B$, locally encode the input states on systems $A$ and $B$ (which may be entangled) such that quantum information can be transferred at positive rates under the channels, and then send the encoded states through the channel from $A$ and $B$ to the middle node $C$, and do the decoding and the unitary $U$ on node $C$, and finally encode the transformed systems $A$ and $B$ and send them through the channels to their original locations, where they are decoded so as to obtain the output of $U$. The encoding and decoding are allowed to be approximate, hence the protocol implements the $U$ approximately.

\bpp\label{prop1}
Let $\cC$ be the class of unitaries of the type \eqref{u:single_group} but with all $W_B(f)$ being invertible. The fast protocol which is the natural analog of Protocol 5 for unitaries in $\cC$ satisfies that the implemented unitary must be of the double-group form \eqref{u:double_group2} up to local unitaries.
\epp
\bpf
If such a generic protocol exists, the final local corrections on $B$ should be unitaries that form a projective representation of a group, similar to Protocols 3 and 5. This fact and the assumption together imply that there is an invertible linear operator $Q$ such that $W_B(f)=c(f) T_B(f) Q$ for any $f\in G$, where $T_B(f)$ are unitaries such that $\{V_A(f)\ox T_B(f)\}_{f\in G}$ is a projective representation of the group $G$, and $c(f)$ are complex numbers that fit the conditions in Protocol 5. Thus $U=\sum_{f\in G} V_A(f)\ox W_B(f)$ is equivalent under a local linear operator $I_A\ox Q$ to another unitary $U'=\sum_{f\in G} c(f) V_A(f)\ox T_B(f)$ acting on the same space. From \cite[Theorem 7]{cy14}, $Q$ must be a unitary. Hence $U$ is equivalent to the form \eqref{u:double_group2} under local unitaries.
\epf

\section{Main results}
\label{sct3}

\subsection{Definitions of the protocols with repeater nodes}\label{ssct:3.1}

The general scenario we consider is the following: there are $n+2$ nodes on a straight line, with the two end nodes denoted $A$ and $B$, and others called repeater nodes. We are allowed to perform local operations and send classical or quantum messages among the nodes. Under the assumptions in Sec.~\ref{sct2}, we aim to perform a unitary on $AB$ with the help of the repeater nodes, or to perform a remote unitary on one end node with parameters determined by the party at the other end node. Apart from those protocols in Sec.~\ref{sct2}, we shall also consider the following variant protocols that involve placing the ancillae in the original protocols at different repeater nodes, and possibly using repeater nodes in more complicated ways. Some of them, such as Protocol 3.3, need more detailed definitions because at least one repeater node is used in ``more complicated ways.'' For those protocols, the refined definitions will appear in Sec.~\ref{ssct:3.2}.

Our naming convention for the protocols is as follows: the name could be Protocol m.n($x_1,x_2,\dots,x_n$) or Protocol m.n.a($x_1,x_2,\dots,x_n$). The ``m'' is the main protocol number, ``n'' is the number of repeater nodes (sometimes called intermediate nodes). The ``a'' (or ``b'' or ``c'') is the minor protocol number if there is more than one protocol with the same numbers m and n. And $x_1,x_2,\dots,x_n$ are the distances of the repeater nodes (which are on the line connecting $A$ and $B$) from $A$ divided by $L$, where $L$ is the distance between $A$ and $B$. We denote the repeater nodes as $C_1, C_2, C_3$, etc, but when there is only one repeater node, we call it $C$.

\begin{figure*}[ht]
\begin{center}
\includegraphics[scale=0.77]{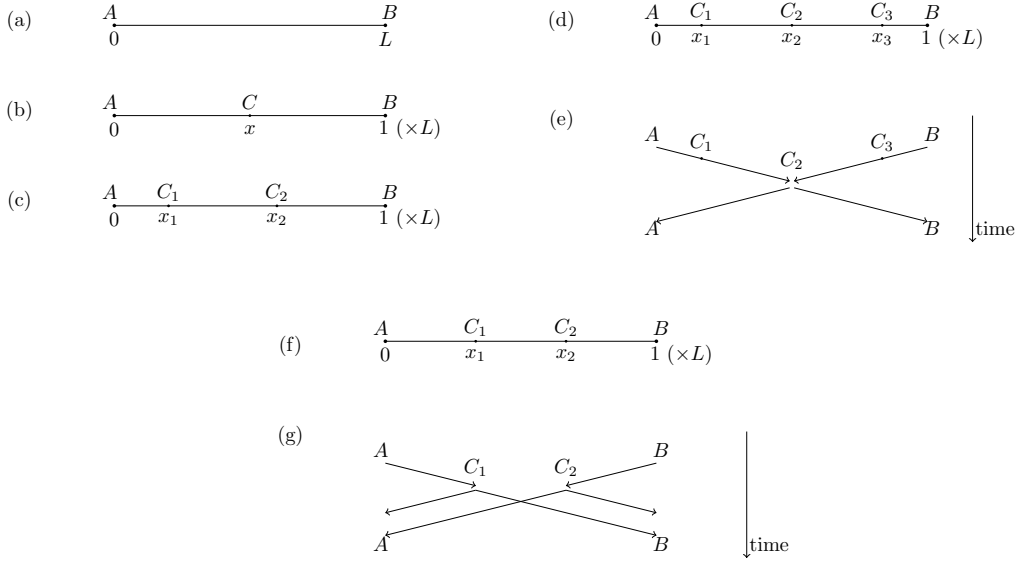}
\end{center}
\caption{The spatial setting of the protocols. The locations of the repeater nodes illustrate the best protocols listed in Table~\ref{tbl_totaltime} in Sec.~\ref{ssct:3.3}. (a) With no repeater node. The distance between $A$ and $B$ is $L$. (b) With one repeater node at position $x L$ (``position'' means distance to $A$). Here $x=\frac{1}{2}$ illustrates Protocol 1.1.  (c) With two repeater nodes at positions $x_1 L,x_2 L$. Here $x_1=\frac{1}{5}$ and $x_2=\frac{3}{5}$, illustrating Protocol 1.2($\frac{1}{5},\frac{3}{5}$). (d) With three repeater nodes at positions $x_1 L,x_2 L,x_3 L$.  Here $x_1=\frac{1}{6}$,  $x_2=\frac{1}{2}$, and $x_3=\frac{5}{6}$, illustrating Protocol 1.3($\frac{1}{6},\frac{1}{2},\frac{5}{6}$). The figure (e) shows the information flow in Protocol 1.3($\frac{1}{6},\frac{1}{2},\frac{5}{6}$). The first step in the protocol which is not shown here is generating entanglement on $A C_1$ (short for ``between $A$ and $C_1$''), and on $B C_2$. Then the input states at $A$ and $B$ are teleported in two steps to $C_2$, with the $C_1$ and $C_3$ acting as relays; the entanglement on $C_1 C_2$ and on $C_3 C_2$ are prepared while the states were being teleported to $C_1$ or $C_3$. Then the target unitary is done locally at $C_2$, and the output systems belonging to $A$ and $B$ are teleported back, with the help of entanglement on $A C_3$ and $B C_3$ prepared in the previous stages. The information flows for the protocols illustrated in (b) and (c) are similar, with the ``middle'' node on which to carry out the target unitary being $C$ in case of (b) and $C_2$ in case of (c). The information flows in variants of Protocol 3 are similar to those of the variants of Protocol 1 mentioned above (but the information being sent is not the whole input state). The figures (f) and (g) are both for Protocol 5.2($\frac{1}{3},\frac{2}{3}$), illustrating the locations of nodes and the flow of information, respectively. In (f), $x_1=\frac{1}{3}$, and $x_2=\frac{2}{3}$. In (g), the information about the computational basis of the ancillary systems $a$ and $b$ are sent to $C_1$ and $C_2$ with the help of entanglement on $A C_1$ and $C_2 B$, and some measurements are done on $C_1$ and $C_2$, and then classical communications are sent to both directions from both $C_1$ and $C_2$, and finally some local unitaries are done on $A$ and $B$ to complete the protocol.}\label{fgr_loc}
\end{figure*}

The general spatial setting of the protocols is illustrated in Fig.~\ref{fgr_loc}, which also shows the information flow in some of the protocols.

{\bf Protocol 1.1.} A node $C$ is at $\frac{L}{2}$ distance to both $A$ and $B$. First perform the usual quantum teleportation from $A$ to $C$ and from $B$ to $C$. The effect is to send the input state of the joint system $AB$ (note the input state could be entangled between $A$ and $B$) to $C$. Then perform the unitary $U$ at $C$, and send the two systems back to their original locations. The spatial setting is illustrated in Fig.~\ref{fgr_loc}(b).

{\bf Protocol 1.2($x_1,x_2$).} The protocol is similar to Protocol 1.1 but has one more repeater node, and thus it contains an extra step of teleporting some quantum state. (The same is generally true for protocols with the same main protocol number but different number of repeaters.) The steps are as follows: Teleport the input state of $A$ to the first repeater node $C_1$ at distance $x_1$ from $A$, and then teleport this state to the second repeater node $C_2$ at $x_2$ from $A$. Meanwhile, teleport the input state of $B$ to $C_2$. Then do the unitary $U$ at the node $C_2$, and send the two systems back to the original locations. The spatial setting with a particular choice of $(x_1,x_2)$ is illustrated in Fig.~\ref{fgr_loc}(c).

{\bf Protocol 1.3($x_1,x_2,x_3$).} Teleport the input state of $A$ to the first repeater node $C_1$ at distance $x_1$ from $A$, and then teleport this state to the second repeater node $C_2$ at $x_2$ from $A$. Meanwhile, teleport the input state of $B$ to the last repeater node $C_3$ at distance $x_3$ from $A$, and then teleport it to the $C_2$. Then do the unitary $U$ at the node $C_2$, and send the two systems back to the original locations. The spatial setting with a particular choice of $(x_1,x_2,x_3)$ is illustrated in Fig.~\ref{fgr_loc}(d).

{\bf Protocol 2.1($x$).} The unitary $U$ is the same as in Protocol 2, but there is an intermediate node $C$ on the line connecting $A$ and $B$ and is at distance $xL$ from $A$. The ancillae $a$ and $b$ are both on the intermediate node $C$. The detailed steps of the protocol are in Sec.~\ref{ssct:3.2}. It turns out that it suffices to consider $x=\frac{1}{2}$ in the current setting, since other values of $x$ are inferior compared to $x=\frac{1}{2}$.

{\bf Protocol 2.1.a($x$).}  Same as the Protocol 2.1($x$) except that the ancilla $a$ for system $A$ is on the intermediate node $C$, while the ancilla $b$ for system $B$ is at the end node $B$.

{\bf Protocol 2.1.b($x$).}  Same as the Protocol 2.1($x$) except that the ancilla $b$ for system $B$ is on the intermediate node $C$, while the ancilla $a$ for system $A$ is at the end node $A$.

In this paper we leave out the detailed description and analysis of the above two protocols, since the best total time we could obtain is not better than that of Protocol 2.1($\frac{1}{2}$). Similarly, for the Protocols 3 through 6, there are similar variants with one repeater node, and we shall also leave out the detailed description and analysis of those protocols, since the best time we could obtain is not better than the case that the ancillary systems $a$ and $b$ are both in the node $C$.

{\bf Protocol 2.2($x_1,x_2$).}  The unitary $U$ is the same as in the Protocol 2, but there are two intermediate nodes $C_1$ and $C_2$, which are on the line $AB$ and are at distances $x_1 L$ and $x_2 L$ from $A$, respectively, where $0<x_1<x_2<1$. The ancillae $a$ and $b$ are located at $C_1$ and $C_2$, respectively. We shall consider all possible choices of $(x_1,x_2)$ and find the choice with the smallest total time cost.

In the modified protocols above, we often prepare entangled states shared between neighboring nodes, and the preparation time for such entanglement is counted in the total time. For the Protocols 3 through 6, some variants analogous to the definitions above can be defined and will be discussed in this paper, and we abbreviate the definitions here. In addition, we may consider the following variant for Protocols 3 through 6. We take Protocol 3 as an example, and the corresponding variants of Protocol 4 through 6 are similarly defined.

{\bf Protocol 3.3($\frac{1}{6},\frac{1}{2},\frac{5}{6}$).} We choose to discuss this special case instead of the general {\bf Protocol 3.3($x_1,x_2,x_3$)}. This protocol is similar to the Protocol 3.1($\frac{1}{2}$) but there are three intermediate nodes $C_1,C_2,C_3$. The ancillae $a$ and $b$ are located at the middle node $C_2$. The $C_1$ and $C_3$ act as relays for sending information from $A$ or $B$ to $C_2$, and are located at $\frac{L}{6}$ distance from the nearest end node.

{\bf Protocols 7.2($x_1,x_2$).} Two repeater nodes $C_1$ and $C_2$ are on the line $AB$ and at distances $x_1 L$ and $x_2 L$ from $A$, respectively.

{\bf Protocol 8.1($x$).} An intermediate node between $A$ and $B$ is added to the settings in Protocol 8. Actually we have not found an exact protocol under this setting that works for the generic choice of the target unitary with less time cost than Protocol 8, but if the angle of rotation in the target unitary is in some special set, or if approximate implementation is allowed, then a protocol is possible, and it will be presented in Sec.~\ref{ssct:3.2}. It suffices to consider the case $x=\frac{1}{2}$ as other choices of $x$ are worse.

{\bf Protocols 8.2($x_1,x_2$) and 8.3($x_1,x_2,x_3$).} Similarly, we consider these protocols only under the restriction that the angle of rotation in the target unitary is in some special set. We shall consider special choices of the repeater locations and remark on their optimality.

In actuality there may be more than three intermediate nodes, but we do not consider that for now, and this is partly justified by the following considerations: Theoretically, adding more repeater nodes would shorten the time needed to send photons to the nearest repeater node, thus reducing the time for entanglement generation, but the reduction in time as a fraction of the total time consumption of the protocol is not very large, while more local gates and measurements associated with the presence of more nodes can introduce more errors.

\subsection{Analysis of the protocols}\label{ssct:3.2}

{\bf Model for entanglement generation.} We use the process in \cite{nemoto14} for generating entanglement between two matter qubits at neighboring nodes $a$ and $b$ (we use small letters to mean generic nodes and to distinguish from the end nodes $A$ and $B$), but we idealize the scheme to temporarily ignore the decoherence in the local systems, while the error in channel transmission is still considered. The process is as follows: The node $a$ generates a photon, and sends it through the circuit shown in Fig.~2(b) of \cite{nemoto14}, so that it may interact with either one of two matter qubits depending on the path it takes. The which-path information is erased because of the way the detector is set up. If there is a detector click, then we know that there is entanglement generated between the matter qubits at $a$ and $b$. It can be regarded as maximally entangled due to the following reasoning. In the current case of photonic channels, the error can be modeled as of two types: the loss error and phase error. The loss error is unimportant because we postselect on the cases when a photon hits the detector shown in Fig.~2(b) of \cite{nemoto14}. The phase error refers to the relative phase between the polarization states in each path that the photon may travel through. But, note that the main (nonlocal) part of the two paths can be actually put in the same optical fiber, which implies that the phase error in the two paths would be almost identical. And it is the difference between the phase errors in the two paths that would affect the final entangled state of the matter qubits, hence the phase error is also not important here. Thus, the scheme in \cite{nemoto14} is resilient to both the photon loss and phase errors. (As stated above, we temporarily ignore the errors from local decoherence effects of the matter qubits; this can be justified by that such error is related to the time of entanglement generation, thus, it is of the same order as other decoherence errors in the whole quantum network in which our target unitary is to be applied to two of the nodes. Such errors will not be ignored if we do a complete error analysis of the whole quantum network after performing one or more unitaries, but for the current purpose of calculating the total time without blowing up the error rate, it can be temporarily ignored.)

Given the considerations about the physical model above, what we will actually use in this paper is the following abstract model: some photons are sent from $a$ to $b$ with possible interactions with matter qubits on the two nodes; some maximally entangled pairs of matter qubits are generated but the entanglement fails to be generated in some other pairs, at the photons' planned arrival time at node $b$; at that particular time, only the party at $b$ knows which matter qubits on $b$ are successfully entangled with some system in $a$, and both parties at $a$ and $b$ know the correspondence relation of the matter qubits on the two nodes, i.e., which qubits are planned to be entangled; then, the party at $b$ sends classical signals to node $a$ so that the party at $a$ knows which matter qubits are entangled with matter qubits on $b$. Since in practice we can only use finitely many identical copies of the above setup to generate entanglement (effectively using many ancillary systems for one input system), the entanglement generation may fail at times. Let us denote by $p$ the probability that all required entangled states are successfully generated in a given nonlocal unitary protocol. By choosing enough copies of local setups, the $p$ can be made to be near $1$ for any fixed finite set of nonlocal unitary protocols on input systems of fixed sizes. The overall operation implemented after one run of the protocol is in general a quantum operation (i.e., a completely-positive trace-preserving map) that approximates the ideal quantum operation corresponding to the target unitary. The error in the approximate implementation would be smaller and smaller if we increase the number of local setups. In the limit of an infinite number of local setups for generating entanglement, the unitary would (ideally) be implemented exactly. In the rest of the paper, the time of protocols or procedures always refers to the time under such idealized scenario.

The process of generating entanglement as mentioned above requires  time
\bea\label{eq:time_ent}
\frac{l}{c}+\frac{l}{c}=\frac{2l}{c},
\eea
where $l$ is the distance between $a$ and $b$, and the $c$ is the speed of light. The first $\frac{l}{c}$ in \eqref{eq:time_ent} is for photon transmission, and the second $\frac{l}{c}$ is for the node $a$ to be notified of the success by means of a classical message from $b$ to $a$. In the protocols below, the second time period of length $\frac{l}{c}$ is often arranged to coincide with other processes in the protocol in order to save time.

{\bf Protocol 1}: The entanglement generation process needs time $\frac{2L}{c}$. It has two parts: first, some photons are sent from $B$ to $A$ and interacts with the matter qubits to generate entanglement, then some classical signals are sent from $A$ to $B$ to indicate which pairs of matter qubits are entangled successfully (this would be called \emph{to confirm entanglement} later). Teleportation of system $A$ to the $B$ side requires time $\frac{L}{c}$, but that could coincide with the above process of confirming entanglement. Then, the target unitary $U$ is performed locally on $B$, and local gates are assumed to take no time. Finally, the teleportation of one system back from $B$ to $A$ with the help of previously established entanglement also requires time $\frac{L}{c}$. The total time is $\frac{3L}{c}$.

{\bf Protocol 1.1}: The first part of the entanglement generation process is sending photons from the middle node $C$ to the end nodes $A$ or $B$, which takes time $\frac{L}{2c}$. The messages for confirming entanglement would be sent along with other messages in the teleportation from $A$ or $B$ to $C$. This teleportation step takes time $\frac{L}{2c}$. The unitary $U$ is done on the $C$. Finally, teleporting back the systems to $A$ and $B$ using prior established entanglement also requires time $\frac{L}{2c}$. The total time is $\frac{3L}{2c}$.

Before we look at the Protocols 1.2($x_1,x_2$) and 1.3($x_1,x_2,x_3$) in general, we first look at their special cases.

{\bf Protocol 1.2($\frac{1}{5},\frac{3}{5}$)}: We first generate the entanglement between $C_1$ and an ancilla on node $A$, and also between $C_2$ and an ancilla on node $B$. The first part of the entanglement generation process is sending of photons from $C_1$ to node $A$, and from $C_2$ to node $B$ (the latter takes longer time). The second part of the entanglement generation is confirming entanglement, which coincides with the sending of classical messages in the teleportation of input states of $A$ to $C_1$, or from $B$ to $C_2$. The $A C_1$ part of the above process takes time $\frac{2L}{5c}$, while on the part on $B C_2$ would take time $\frac{4L}{5c}$ which would partially overlap with the later steps below. The entanglement between $C_1 C_2$ is also generated but not confirmed yet during the first time period of $\frac{2L}{5c}$. Then, we carry out the teleportation of the state at $C_1$ (which is now the same as the original input state of $A$) to $C_2$, together with classical communication from $C_1$ to $C_2$ for confirming entanglement. This takes time $\frac{2L}{5c}$. After that, the state teleported from $B$ would have just arrived at $C_2$. Then the unitary $U$ is carried out locally at $C_2$, and the systems belonging to $A$ and $B$ are teleported back, using entanglement created during previous time periods. This last time period is of length $\frac{3L}{5c}$. The total time needed is $\frac{7L}{5c}$.

{\bf Protocol 1.3($\frac{1}{6},\frac{1}{2},\frac{5}{6}$)}: We first generate the entanglement between $C_1$ and an ancilla on node $A$, and also between $C_3$ and an ancilla on node $B$. The first part of the entanglement generation process is sending of photons from $C_1$ to node $A$, and simultaneously from $C_3$ to node $B$. The second part of the entanglement generation is confirming entanglement, which coincides with the teleportation (of the input states) from $A$ to $C_1$ and simultaneously from $B$ to $C_3$. The above steps take time $\frac{L}{3c}$. The entanglement between $C_1 C_2$ and between $C_2 C_3$ is also generated but not confirmed during the above time period. Then we carry out the teleportation of the states at $C_1$ and $C_3$ to $C_2$, which coincides with confirming entanglement on $C_1 C_2$ and $C_2 C_3$. This takes time $\frac{L}{3c}$. Then the unitary $U$ is carried out locally at $C_2$, and the systems belonging to $A$ and $B$ are teleported back, using entanglement created during previous time periods. This last time period is of length $\frac{L}{2c}$. The total time needed is $\frac{7L}{6c}$. The location of nodes and the flow of information of this protocol are illustrated in Figs.~\ref{fgr_loc}(d) and \ref{fgr_loc}(e).

As mentioned in Sec.~\ref{sct1}, the information flow in this protocol is typical of the types of protocols that we consider, and thus we review it here. The information about the input system at each one of the two end nodes is teleported in several steps to a ``middle'' node (it is exactly at equal distances to the two ends in this example, but generally does not have to be the middle one either in terms of spatial locations or in terms of number of nodes) through an array of repeaters with the help of pairwise entanglement between neighboring nodes (which is prepared in an early stage of the protocol or sometimes in parallel with some other steps in the protocol), then the target unitary is performed at the ``middle'' node, and then the states belonging to each of the two output systems are teleported to the respective end node. Thus, we see that only the one repeater located at the ``middle'' needs to perform more complex operations than what the usual repeaters do in preparing long-range entanglement.

{\bf Protocols 1.2($x_1,x_2$) and 1.3($x_1,x_2,x_3$) with generic choices of $x_1,x_2,x_3$}: If the steps of the protocols are similar to the special cases above, the time cost of the generic cases is the same as the corresponding variants of Protocol 3.  When the operators $V_A(f)$ and $T_B(f)$ defined in Protocol 3 are chosen as generalized Pauli operators and the group $G$ chosen to be of size $d_A^2 d_B^2$, not only is the total time cost the same, but the total entanglement cost (required number of photons) differ by at most a factor of four (with the cost in variants of Protocol 3 larger than that in the variants of Protocol 1). The Protocol 3 is more general in the sense that the group $G$ is not limited to the generalized Pauli group, and for special classes of unitaries $U$ with sufficiently small $G$, the entanglement cost of the variants of Protocol 3 could be smaller than the corresponding variants of Protocol 1 by an arbitrarily large factor. We leave out the analysis of time cost with the generic choices of repeater positions here, since that is completely isomorphic with the corresponding cases for variants of Protocol 3. Protocol 1.2($\frac{1}{5},\frac{3}{5}$) is optimal for two repeater nodes, for reasons similar to the analysis later for Protocol 3.2($x_1,x_2$). In the case of three repeater nodes, Protocol 1.3($\frac{1}{6},\frac{1}{2},\frac{5}{6}$) is optimal among the protocols with similar steps but different repeater locations, but there might be a protocol satisfying the same constraints but with entirely different internal mechanism and less total time cost.

\begin{figure*}[ht]
\begin{center}
\includegraphics[scale=1]{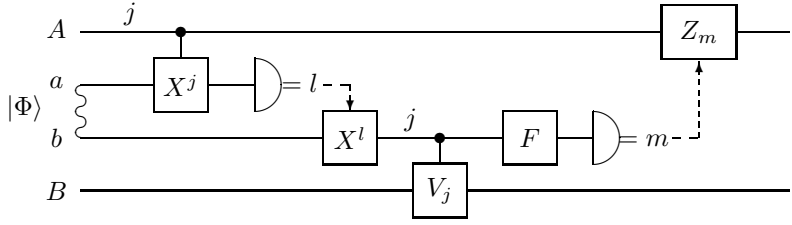}
\end{center}
\caption{The circuit diagram from \cite{ygc10} for implementing a bipartite controlled unitary on $AB$ of the form \eqref{u:control}. In the variants of Protocol 2 and also some other protocols in this paper, this circuit serves as a primitive, with the $A$ and $B$ in the figure representing the two systems where a controlled unitary (not necessarily the target unitary, but could be some auxiliary gate in a protocol) is performed on.} \label{fgr_cont}
\end{figure*}

{\bf Protocol 2}: We show the circuit in Fig.~\ref{fgr_cont} which is from \cite{ygc10}. This is partly for the current analysis, but also partly for helping the analysis in the variants of Protocol 2 and even those of Protocols 3 through 6 below. Assume the ancillae $a$ and $b$ are (nearly) at the locations of the input systems $A$ and $B$, respectively. The first part of the entanglement generation process (sending photons from party $B$ to party $A$) takes time $\frac{L}{c}$, and the second part which is confirming entanglement is in parallel with the sending of classical messages from party $A$ to party $B$ in the controlled unitary protocol, which takes time $\frac{L}{c}$. The last part of the protocol, which involves sending a classical message from party $B$ to party $A$, also takes time $\frac{L}{c}$. The total time is $\frac{3L}{c}$.

{\bf Protocol 2.1($\frac{1}{2}$)}: The protocol is based on Protocol 2. Entanglement is to be generated for doing the controlled-$X^j$ gate on $Aa$. The first part of the entanglement generation process (sending photons from the middle node $C$ to $A$) takes time $\frac{L}{2c}$. The generated entangled state is on two extra ancillary systems different from $A$ and $a$. The second part of the entanglement generation process coincides with the sending of classical messages from $A$ to $C$ in the protocol for performing the controlled-$X^j$ gate (with $A$ being the control), which takes time $\frac{L}{2c}$. The last part of the protocol for performing the controlled-$X^j$ gate on $Aa$ also takes time $\frac{L}{2c}$. This last step could coincide with the first half of the time period to perform the controlled-$V_j$ gate on $bB$. The entanglement generation between $bB$ could be done in parallel with the previous steps. The controlled-$V_j$ gate on $bB$ using prior established entanglement consumes time $2\frac{L}{2c}=\frac{L}{c}$. Then, the $b$ is measured, and the classical outcome is sent to $A$, which takes time $\frac{L}{2c}$. The total time is $\frac{5L}{2c}$.

{\bf Protocol 2.2($x_1,x_2$)}: The analysis is in Appendix~\ref{app:ptl32}. It is shown there that the choice of $(x_1,x_2)$ with the shortest total time is $(\frac{1}{7}, \frac{3}{7})$.  In the following we describe the protocol in this special case.

{\bf Protocol 2.2($\frac{1}{7}, \frac{3}{7}$)}: Entanglement needs to be generated on $AC_1$ in order to carry out the controlled-$X^j$ gate on $AC_1$. The first part of the entanglement generation process (sending photons from $C_1$ to $A$) takes time $\frac{L}{7c}$. The second part (to confirm entanglement by sending classical messages from $A$ to $C_1$) coincides with the sending of classical message from $A$ to $C_1$ in the controlled unitary protocol on $A C_1$, which takes time $\frac{L}{7c}$. The first part of entanglement generation between $C_1$ and $C_2$ takes time $\frac{2L}{7c}$, thus it could coincide with the time periods of the two previous steps. After measurement on $C_1$, the sending of the measurement outcome from $C_1$ to $C_2$, together with the second part of the entanglement generation on $C_1 C_2$, takes time $\frac{2L}{7c}$. The first part of the entanglement generation on $C_2 B$ takes time $\frac{3L}{7c}$, which could be contained in the time interval for above steps. The second part (to confirm entanglement) could be contained in the time interval for the protocol of implementing the controlled-$V_j$ gate on $C_2 B$ using entanglement on $C_2 B$. Such protocol on $C_2 B$ consumes time $2\frac{4L}{7c}=\frac{8L}{7c}$. Then, the $C_2$ is measured, and the classical outcome is sent to $A$, which takes time $\frac{3L}{7c}$. The total time is $\frac{2L}{7c} + \frac{2L}{7c} + \frac{8L}{7c} + \frac{3L}{7c} = \frac{15L}{7c}$.

Thus, among the variants of Protocol 2 with two repeater nodes, the Protocol 2.2($\frac{1}{7},\frac{3}{7}$) has the smallest time cost $\frac{15L}{7c}$. On the other hand, if there is only one repeater node, the Protocol 2.1($\frac{1}{2}$) has a time cost of $\frac{5L}{2c}$, larger than the $\frac{3L}{2c}$ for Protocol 1.1, but the entanglement cost (the required number of photons) here could be much less than that of Protocol 1.1 for the same controlled unitary $U$, when the projectors $P_j$ in \eqref{u:control} are of high rank, or when $d_B$ is large.\\

\begin{figure*}[ht]
\begin{center}
\includegraphics[scale=1]{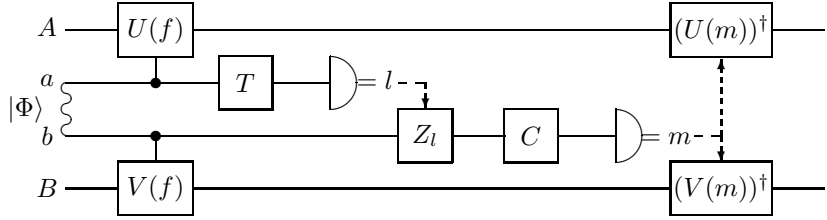}
\end{center}
\caption{The circuit diagram for implementing a bipartite unitary on $AB$ of the form \eqref{u:double_group}. It is used for Protocol 3 (and thus illustrates the main structure of the variants of Protocol 3). The gate $C$ is denoted as $\hat C$ in the text to distinguish from the node $C$, such as in the analysis of Protocol 3.1($\frac{1}{2}$).} \label{fgr_group}
\end{figure*}

{\bf Protocol 3}: Since this protocol is to be reused many times in this paper, and the paper \cite{ygc10} only shows a reduced version of the circuit, we show the complete circuit in Fig.~\ref{fgr_group}, which also appears in \cite{ygc12}.

We consider the generic case, that is, not using the fast protocols in \cite{ygc12} which works for some of the unitaries of the current form. The first part of the entanglement generation process (sending photons from $B$ to $A$) takes time $\frac{L}{c}$, and the second part coincides with the sending of classical messages from $A$ to $B$, which takes time $\frac{L}{c}$. The last part of the protocol, which involves sending a classical message from $B$ to $A$, also takes time $\frac{L}{c}$.  The total time needed by the protocol is $\frac{3L}{c}$.

{\bf Protocol 3.1($\frac{1}{2}$)}: The local systems $a$ and $b$ at the middle node $C$ are initially in a maximally entangled state as required by Protocol 3. The entanglement between the middle node $C$ and the two end nodes $A$, $B$ are created in order to perform the controlled-$V_A(f)$ gate on $aA$ and the controlled-$T_B(f)$ gate on $bB$ using the usual protocol for Protocol 4, where $a$ and $b$ are the controls. Denote the entangled systems as $ef$ and $gh$, where $e$ is on $A$, $f$ and $g$ are on $C$, and $h$ is on $B$. The reason why Protocol 4 is good for implementing the controlled-$V_A(f)$ [and controlled-$T_B(f)$] gate is because the $\{V_A(f)\}$ [and the $\{T_B(f)\}$] is a projective representation of a group. The first part of the process of creating such entanglement is by sending photons from the middle node to the end nodes, and it takes time $\frac{L}{2c}$. The second part is confirming entanglement, which coincides with the sending of classical messages from the end nodes to the middle node in the usual protocol for Protocol 4, and takes time $\frac{L}{2c}$. The second part of the usual protocol for Protocol 4 involves sending classical messages from the middle node to the end nodes, and this could coincide with the last step of the main protocol which also sends classical messages from the middle node to the end nodes, after suitable local operations on $a$ and $b$ are done. The local operations involve measuring the $a$ in the Fourier basis and performing corresponding phase corrections on $b$, and performing a $\hat C$ gate (refers to the group circulant matrix $C$ in the protocol for the double-group type of protocol in \cite{ygc10}, same below) on $b$, and measuring $b$ in the standard basis. The time cost for the local operations is ignored, and the last step of communication takes time $\frac{L}{2c}$. Finally, some local gates are performed on $A$ and $B$ according to the received classical messages from the middle node. The total time needed is $\frac{3L}{2c}$.

In Protocol 3.1($\frac{1}{2}$), the parameters $c(f)$ in the expression of $U$ only enter through the operations at the middle node, but the operations on $A$ and $B$ still depend on the group $G$ and the forms of $V_A(f)$ and $T_B(f)$. In this sense, the unitary $U$ is to some degree remotely determined by the middle party. This has some resemblance to the remote single-qubit unitary of Protocol 8. One might even say that the Protocol 3.1($\frac{1}{2}$) implements a more complicated version of a remote unitary: the unitary is on two parties and determined by a third party on the middle node, while the unitary in Protocol 8 is a local unitary determined by another party.

{\bf Protocol 3.2($x_1,x_2$)}: Since there are a few different cases, we illustrate the cases by some particular choices of $(x_1,x_2)$ first, and then consider the generic case.

{\bf Protocol 3.2($\frac{1}{3},\frac{2}{3}$)}: We first generate the entanglement between an ancilla called $a'$ on node $C_1$ and an ancilla on node $A$, and also between an ancilla called $b'$ on node $C_2$ and an ancilla on node $B$. The first part of the entanglement generation process is sending of photons from node $C_1$ to node $A$, and from node $C_2$ to node $B$. The second part of the entanglement generation process coincides with the sending of classical messages (from $A$ to $C_1$, and from $B$ to $C_2$) in the first communication step in the protocol for the controlled-$V_A(f)$ [or controlled-$T_B(f)$] gate, which uses the usual protocol for Protocol 4. The above steps take time $\frac{2L}{3c}$. Let $a$ and $b$ be some other ancillae on $C_1$ and $C_2$, respectively. The entanglement between $a$ and $b$ is also generated during the above time period. Then, we continue to implement the controlled-$V_A(f)$  [controlled-$T_B(f)$] gate on $a$ and $A$ ($b$ and $B$), by doing suitable local operations on node $C_1$ ($C_2$). Then some local measurement is performed on $a$, with the outcome sent classically to node $b$, taking time $\frac{L}{3c}$. In this time period the last communication step of the protocol for the controlled-$V_A(f)$ [or controlled-$T_B(f)$] gate, i.e. sending messages to $A$ ($B$) is finished. Then a local correction is done on $b$ according to the received message from $a$, and the gate $\hat C$ is done on $b$, followed by a computational basis measurement of $b$, and the outcome is sent to both $A$ and $B$, taking time $\frac{2L}{3c}$. The protocol is completed by doing local unitary corrections at $A$ and $B$. The total time needed is $\frac{5L}{3c}$.

{\bf Protocol 3.2($\frac{1}{5},\frac{3}{5}$)}: The steps are similar to those in Protocol 3.2($\frac{1}{3},\frac{2}{3}$). The first and second time periods are both of length $\frac{2L}{5c}$. And during this combined time period of $\frac{4L}{5c}$, the entangled state between $b'$ at $C_2$ and an ancilla at $B$ is established, and the first half of the procedure for performing the controlled-$T_B(f)$ gate using the usual protocol for Protocol 4 is completed. Some local measurement is performed on $b$ and $b'$ which are both at node $C_2$. Then the outcome of the measurement on $b'$ is sent to $B$ in parallel with the sending of the measurement outcome of $b$ to the two end nodes. The final time period is of length $\frac{3L}{5c}$, since the $b$ is $\frac{3L}{5}$ from $A$ and only $\frac{2L}{5}$ from $B$. The total time needed is $\frac{7L}{5c}$.

{\bf Generic choices of $(x_1,x_2)$ for Protocol 3.2($x_1,x_2$).} The analysis is in Appendix~\ref{app:ptl32}, since this part is long and only serves to show why Protocol 3.2($\frac{1}{5},\frac{3}{5}$) has the smallest total time cost among all choices of $(x_1,x_2)$ under our type of protocols.

{\bf Protocol 3.3($\frac{1}{6},\frac{1}{2},\frac{5}{6}$)}: This protocol is very much like Protocol 3.1($\frac{1}{2}$). The first and last intermediate nodes $C_1$ and $C_3$ act as relays to reduce the total time consumption. Initially, the systems $a$ and $b$ on $C_2$ are locally maximally entangled according to the requirement in the Protocol 3.
To establish entanglement between neighboring nodes, some photons are sent from $C_1$ to $A$ and from $C_2$ to $C_1$, and also from $C_2$ to $C_3$ and from $C_3$ to $B$.  Then some classical messages are sent from $A$ to $C_1$, and from $B$ to $C_3$, to indicate which atoms had been entangled, as well as to send the classical messages required in the protocol for implementing the controlled-$V_A(f)$ [controlled-$T_B(f)$] gate where the control is by system $a$ at $C_2$ (system $b$ at $C_2$). The above steps take time $\frac{L}{3c}$. At the end of this time period, some photons from $C_2$ would have arrived at both $C_1$ and $C_3$. Then the state of $e$ ($f$) is teleported to $C_2$ using the entanglement between $C_1 C_2$ ($C_2 C_3$), together with classical messages indicating which atoms had been entangled. This takes time $\frac{L}{3c}$. Denote the teleported system for $e$ and $f$ as $e'$ and $f'$, respectively. At the middle node $C_2$, some local operations for the controlled-$V_A(f)$ [controlled-$T_B(f)$] gate are done, with measurement outcomes to be sent to the relevant end node ($A$ or $B$) together with the measurement outcome of $b$ described below. Then local operations on $C_2$ according to some steps in the original Protocol 3 are performed: the $a$ is measured in the Fourier basis, with a phase correction on $b$, and a $\hat C$ gate is performed on $b$, and then $b$ is measured in the computational basis, and the results are sent to $A$ and $B$. The sending of measurement outcomes on $b$, $e'$, and $f'$ to the end nodes takes time $\frac{L}{2c}$. Finally local unitary corrections are performed on $A$ and $B$. The total time needed is $\frac{L}{3c}+\frac{L}{3c}+\frac{L}{2c}=\frac{7L}{6c}$. The location of nodes and the flow of information of this protocol are illustrated in Figs.~\ref{fgr_loc}(d) and \ref{fgr_loc}(e). The same figures also illustrate Protocol 1.3($\frac{1}{6},\frac{1}{2},\frac{5}{6}$). The only differences are in the local operations (including the use of local ancillas), and the type of information being sent: in Protocol 1.3($\frac{1}{6},\frac{1}{2},\frac{5}{6}$), the whole quantum state was being sent through teleportation, but here the information being sent is some particular type of information about the input quantum state.

Therefore, among the above variants of Protocol 3 with up to three repeater nodes, the Protocol 3.3($\frac{1}{6},\frac{1}{2},\frac{5}{6}$) has the smallest time cost, which is $\frac{7L}{6c}$.
Among the variants of Protocol 3 with at most two repeater nodes, the Protocol 3.2($\frac{1}{5},\frac{3}{5}$) has the smallest time cost which is $\frac{7L}{5c}$. When there is only one repeater node, the Protocol 3.1($\frac{1}{2}$) has a time cost of $\frac{3L}{2c}$. This can be compared to $\frac{5L}{2c}$ for the variants of Protocol 2 with one repeater node. The difference arises because in \eqref{u:control}, the $V_j$ do not generally form a (projective) representation of a group. The latter is also the reason we do not have a variant of Protocol 2 that is similar to Protocol 3.3($\frac{1}{6},\frac{1}{2},\frac{5}{6}$).\\

\begin{figure*}[ht]
\begin{center}
\includegraphics[scale=0.9]{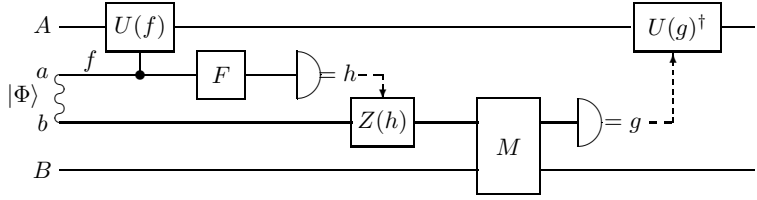}
\end{center}
\caption{The circuit diagram from \cite{ygc10} for implementing a bipartite unitary on $AB$ of the single-group form \eqref{u:single_group}. It is used for Protocol 4.} \label{fgr_gengroup}
\end{figure*}

{\bf Protocol 4}: We consider the generic case, that is, not using Protocol 5 which works for some of the unitaries of the current form. The protocol is illustrated in Fig.~\ref{fgr_gengroup}.  The systems $a$ and $A$ are both located at the end node $A$, and the systems $b$ and $B$ at the end node $B$, hence the time needed by the protocol is $\frac{3L}{c}$, by using the same operating sequence as Protocols 2, just with the local gates and measurements changed.

The variants of Protocol 4 can also be carried out using similar procedures as those for the variants of Protocol 2, but there are a few differences: The first controlled gate is controlled from $a$ rather than $A$, and this does not affect the time for completing such gate. The entanglement generation process on $aA$ is such that photons are sent from $A$ to $a$ first, before the classical signals for confirming entanglement; and then the basic controlled unitary protocol (Protocol 2) is used to implement the controlled gate on $aA$ with $a$ as control, so the directions of the two stages of classical communication are opposite to those in the corresponding variant of Protocol 2. The $M$ gate is to be carried out using Protocol 1, using the same amount of time as the corresponding controlled gate on $bB$ in Fig.~\ref{fgr_cont}. The total time needed is the same as the corresponding variant for Protocol 2.

Thus, among the variants of Protocol 4 with up to two repeater nodes, the Protocol 4.2($\frac{1}{7},\frac{3}{7}$) has the smallest time cost, which is $\frac{15L}{7c}$.

\begin{figure*}[ht]
\begin{center}
\includegraphics[scale=1]{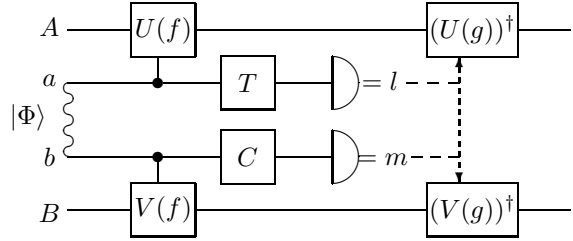}
\end{center}
\caption{The circuit diagram from \cite{ygc12} for implementing a bipartite unitary on $AB$ of the double-group form \eqref{u:double_group} using the fast double-group protocol (Protocol 5).} \label{fgr_fastgroup}
\end{figure*}

{\bf Protocol 5}: If the Protocol 3 is used to implement $U$, it takes time $\frac{3L}{c}$. Alternatively, we can prepare the entangled state without any repeaters, and it takes time $\frac{2L}{c}$ (including the time to transmit photons in one direction and sending classical signals in the other direction), and then the fast protocol shown in Fig.~\ref{fgr_fastgroup} needs time $\frac{L}{c}$. So the total time is $\frac{3L}{c}$ for both methods.

{\bf Protocol 5.1($\frac{1}{2}$)}: By using Protocol 3.1($\frac{1}{2}$), the time is $\frac{3L}{2c}$. It appears that the fact that $U$ fits the fast unitary form is not useful here.

{\bf Protocol 5.2($\frac{1}{3},\frac{2}{3}$)}: The first steps are the same as in Protocol 3.2($\frac{1}{3},\frac{2}{3}$), up to the time $\frac{2L}{3c}$. The entanglement between $a$ and $b$ [as defined in Protocol 3.2($\frac{1}{3},\frac{2}{3}$)] is prepared by then. Some local operations are done on $C_1$ and $C_2$, including the gate $\hat C$ on $b$. Then the $a$ and $b$ are measured in some suitable local bases simultaneously, and the outcomes are sent classically to both end nodes $A$ and $B$. During this last time period of $\frac{2L}{3c}$, the first half of it coincides with the last part of the message sending from $a$ to $A$ and from $b$ to $B$ in the procedure for implementing the controlled-$V_A(f)$ [or controlled-$T_B(f)$] gate. Some local unitary corrections on $A$ and $B$ are performed at the end. The total time needed is $\frac{4L}{3c}$. This means that for the same unitary, the current protocol is better than Protocols 1.2($\frac{1}{5},\frac{3}{5}$) and 3.2($\frac{1}{5},\frac{3}{5}$) in terms of time cost, but not necessarily in terms of entanglement cost, as the latter depends on how large the group $G$ is compared to the size of the input systems. The location of nodes and the flow of information of this protocol are illustrated in Fig.~1 (f)(g).

{\bf Protocol 5.2($\frac{1}{5},\frac{3}{5}$)}: The steps in Protocol 3.2($\frac{1}{5},\frac{3}{5}$) gives total time $\frac{7L}{5c}$. It appears that the fact that $U$ fits the fast unitary form is not useful here.

{\bf Protocol 5.3($\frac{1}{6},\frac{1}{2},\frac{5}{6}$)}: The procedure is the same as that in Protocol 3.3($\frac{1}{6},\frac{1}{2},\frac{5}{6}$), and the total time is $\frac{7L}{6c}$.

Thus, among the above variants of Protocol 5, the Protocol 5.3($\frac{1}{6},\frac{1}{2},\frac{5}{6}$) has the smallest total time cost, which is $\frac{7L}{6c}$. But among the protocols which use at most two repeater nodes, Protocol 5.2($\frac{1}{3},\frac{2}{3}$) is the best which takes time $\frac{4L}{3c}$.

\begin{figure*}[ht]
\begin{center}
\includegraphics[scale=1]{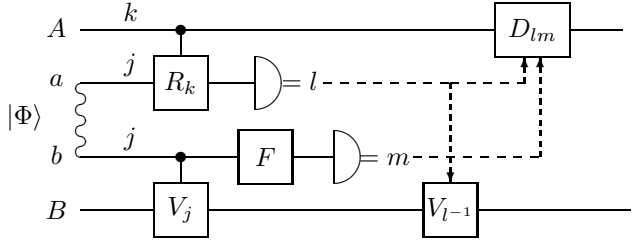}
\end{center}
\caption{The circuit diagram from \cite{Yu11} for implementing a controlled-group unitary $U$ on $AB$. The unitary $U$ is of the form \eqref{u:control} but with the $V_j$ forming a subset of a projective representation of a finite group. It is used for Protocol 6.} \label{fgr_ctgroup}
\end{figure*}

{\bf Protocol 6}: As mentioned previously, in the case of fast controlled-Abelian group protocol in \cite[Sec.~II]{ygc12}, the $U$ is equivalent under local unitaries to the form of \eqref{u:double_group2}, thus can be implemented using the Protocol 5. The best variant with the smallest total time and only two repeater nodes is Protocol 5.2($\frac{1}{3},\frac{2}{3}$) which has total time cost $\frac{4L}{3c}$. For the more general case of controlled-group unitary, we may use a protocol called Protocol 6.2($\frac{1}{3},\frac{2}{3}$) which is similar to Protocol 5.2($\frac{1}{3},\frac{2}{3}$), and is revised from the latter following the controlled-group protocol in \cite{Yu11}. [Among other changes, the controlled-$V_A(f)$ gate on $aA$ is replaced by the controlled-cyclic-shift gate on $Aa$, where $A$ is the control.]  This protocol also takes total time $\frac{4L}{3c}$.
The controlled-group protocol in \cite{Yu11} is shown in Fig.~\ref{fgr_ctgroup}.

Thus, in general the controlled-group unitaries can be implemented in total time $\frac{4L}{3c}$ with two repeater nodes.
The Protocol 6.2($\frac{1}{3},\frac{2}{3}$) should fit the same figures as Protocol 5.2($\frac{1}{3},\frac{2}{3}$) does in Figs.~\ref{fgr_loc}(f) and \ref{fgr_loc}(g), but the interpretation of the flow of information is somewhat  different. With three repeater nodes, a modified version of Protocol 5.3($\frac{1}{6},\frac{1}{2},\frac{5}{6}$) could be used, with total time cost $\frac{7L}{6c}$. The modification follows from the controlled-group protocol in \cite{Yu11}.

{\bf Protocol 7}: Firstly prepare a maximally entangled state without any repeaters, and it takes time $\frac{2L}{c}$ (including the time to transmit photons in one direction and sending classical signals in the other direction for confirming entanglement), and then the main steps of the fast protocol needs time $\frac{L}{c}$. So the total time cost is $\frac{3L}{c}$.

{\bf Protocol 7.2($\frac{1}{3},\frac{2}{3}$)}: Firstly, the entanglement between $A$ and $C_1$, and between $B$ and $C_2$ are prepared. The time for transmitting photons from $C_1$ to $A$ (from $C_2$ to $B$) takes time $\frac{L}{3c}$. The step for confirming entanglement is then done in parallel with teleporting the input state on $A$ (respectively, $B$) to $C_1$ (respectively, $C_2$). This also takes time $\frac{L}{3c}$. By then, the total time from the start is $\frac{2L}{3c}$, so the entanglement between $C_1$ and $C_2$ could be established and confirmed. Then the original type of fast protocol in Protocol 7 is used to implement the unitary in the fast manner on $C_1 C_2$, which takes time $\frac{L}{3c}$. Then the output systems are teleported back to $A$ and $B$ in time $\frac{L}{3c}$. The total time cost is $\frac{4L}{3c}$. If the original protocol without repeaters implements $U$ approximately, then the current variant is also an approximate protocol. It can be determined that the choice of the locations for the two repeaters here is optimal. We abbreviate the detailed reasoning since it is similar to the analysis for other protocols with two repeater nodes discussed previously.

{\bf Protocol 8}: The original protocol in \cite[Fig.~3]{hpv02} requires 1 ebit of entanglement and also the sending of classical information in two directions which are not at the same time, using total time $\frac{3L}{c}$ if the time for preparing entanglement is counted in (which partially overlaps with the later part of the protocol).

{\bf Protocol 8.1($\frac{1}{2}$)}: It is unknown if using one repeater node at the middle would improve the time cost compared to Protocol 8, when the angle of rotation in the target unitary is unrestricted. But when the rotation angle $\theta$ in the target unitary $U=\diag(e^{i\theta},e^{-i\theta})$ is $\pi/2^N$ for some positive integer $N$, we may use the following method to exactly implement $U$, which acts on Bob's input data qubit. It works similarly with $\theta=q\pi/2^N$, where $q$ is an odd integer. The protocol satisfies that the information about the $\theta$ is only in the initial states prepared by Alice and not in the local gates performed after the protocol begins. The steps are as follows:\\
1. Alice locally prepares $N$ two-qubit maximally entangled states of the form $\frac{1}{\sqrt{2}}(e^{i2^{k-1}\theta}\ket{00}+e^{-i 2^{k-1}\theta}\ket{11})$, where $k=1,2,\dots,N$. In the particular case $q=1$, it is $\frac{1}{\sqrt{2}}(e^{i\pi/2^{N-k+1}}\ket{00}+e^{-i\pi/2^{N-k+1}}\ket{11})$. We ignore the amount of time for preparing such state.\\
2. Some photons are sent from the middle node $C$ to $A$ and $B$ for the purpose of establishing entanglement between $AC$ and between $BC$. This takes time $\frac{L}{2c}$.\\
3. Alice teleports the $N$ two-qubit states to $C$ using established entanglement between $AC$, together with some classical signals to $C$ indicating which of $C$'s qubits had been entangled to Alice's (now consumed) qubits successfully. At the same time, Bob does some local operations including a measurement and sends the measurement outcome to $C$, as in the circuit in \cite[Fig.~3]{hpv02}. Along with this message are some classical signals to $C$ indicating which of $C$'s qubits had been entangled with Bob's qubits successfully (some of the entanglement is consumed by now but some is to be used for step 5). This takes time $\frac{L}{2c}$. Now some type of (but not all) information about Bob's input data qubit is on the node $C$.\\
4. The gate $U_{com}$ in \cite[Fig.~3]{hpv02}, which is $U=\diag(e^{i\theta},e^{-i\theta})$ in our setting, is carried out using the following means: it may have up to $N$ steps, but may terminate early depending on the intermediate measurement results. This procedure can be viewed as the single-party version of that in \cite{cdk01}. At the $k$-th step ($1\le k\le N$), we apply a circuit similar to the usual teleportation circuit, but with the entangled state being the $k$-th two qubit state prepared in step (1). This circuit in the $k$-th step can also be viewed as a modified version of the gate teleportation circuit in \cite[Fig. 4]{gc99}, with the entangled state of the specific form as described above, but with the final corrections changed to the usual corrections in teleportation ($X^j Z^{j'}$ up to a global phase, where $j,j'\in\{0,1\}$). Depending on the outcome of the measurements, one of the two single-qubit gates is carried out with probability $\frac{1}{2}$ each: $V_k=\diag(e^{i 2^{k-1}\theta},e^{-i 2^{k-1}\theta})$ or $V'_k=V_k^\dag$. If $V_1$ were carried out, the process terminates. Otherwise, $V'_1$ were carried out on the local qubit on $C$, and we need to apply $V_2$ to correct for that, but if the $V'_2$ is actually implemented in the second step, we then need to continue to do step 3, and again it succeeds with probability $\frac{1}{2}$, and in case of failure we proceed to the fourth step, and so on. At the $N$-th step, $V_{N}=\diag(e^{i q\pi/2},e^{-i q\pi/2})$ and it differs from $V'_{N}$ only by a global phase $(-1)$, and such phase can be corrected if needed. Thus the desired gate corresponding to $U_{com}$ in \cite[Fig.~3]{hpv02} is always implemented successfully. We ignore the time for doing these local operations on the middle node $C$.\\
5. The remaining part of the circuit in \cite[Fig.~3]{hpv02} (the part in the $G_2$ box) is carried out on the parties $C$ and $B$, using entanglement between $C$ and $B$ prepared previously, and the output single qubit state is on $B$. This takes time $\frac{L}{2c}$.

The overall time needed is $\frac{3L}{2c}$, which is half of that needed by Protocol 8. For other values of $\theta$, we may approximate it by $q\pi/2^N$ for some integers $N$ and $q$, but then the implementation of the unitary would be approximate.

{\bf Protocol 8.2($\frac{1}{5},\frac{3}{5}$)}: Again, the following procedure implements $U$ exactly when $\theta=q\pi/2^N$ where $N$ is a positive integer, and $q$ is an odd integer, and it implements $U$ approximately for other values of $\theta$. The states containing the information about $\theta$ are sent to the second repeater node $C_2$ with the help of the first repeater node $C_1$ acting as a relay. Some information about the state on the input $B$ is sent to $C_2$ in a similar way as in Protocol 8.1($\frac{1}{2}$), after entanglement is prepared between $C_2$ and $B$. And after some local operations on $C_2$, the information is sent back to $B$ again using entanglement between $C_2$ and $B$. The timings are similar to those in Protocols 1.2($\frac{1}{5},\frac{3}{5}$) and 3.2($\frac{1}{5},\frac{3}{5}$). The total time needed is $\frac{7L}{5c}$. This time cost is optimal among different choices of repeater positions $(x_1,x_2)$ when the protocol uses the above steps.

{\bf Protocol 8.3($\frac{1}{6},\frac{1}{2},\frac{5}{6}$)}: Similarly, the following procedure implements $U$ exactly when $\theta=q\pi/2^N$ where $N$ is a positive integer, and $q$ is an odd integer, and it implements $U$ approximately for other values of $\theta$. The states containing the information about $\theta$ are sent to the node $C_2$ with the help of the node $C_1$ acting as a relay. Some information about the state on the input $B$ is sent to $C_2$ with the help of $C_3$ as a relay. And after some local operations on $C_2$, the information is sent back to $B$ again using entanglement between $C_2$ and $B$. The timings are similar to those in Protocols 1.3($\frac{1}{6},\frac{1}{2},\frac{5}{6}$) and 3.3($\frac{1}{6},\frac{1}{2},\frac{5}{6}$). The total time needed is $\frac{7L}{6c}$. This time cost is optimal among different choices of repeater positions $(x_1,x_2,x_3)$ when the protocol uses the above steps. But we do not rule out the possibility that the task could be carried out through a different procedure with less total time cost.

\subsection{The minimum total time with given number of repeater nodes}\label{ssct:3.3}

The results of the previous subsection, together with some extra discussion of the simple case of no repeater nodes, and the cases of approximate implementation, are summarized in the following four points.

(i). For implementing a general unitary $U$ on $AB$, without using any intermediate nodes, the total time needed is $\frac{3L}{c}$ under Protocol 1. There appears to be no other exact implementation scheme with a shorter total time. On the other hand, if approximate implementation is allowed, we may first locally encode the input state of one system (say $A$) such that the code is resilient to loss errors, and then send the encoded state through the channel, and do the decoding and the unitary $U$ on the receiving party, and finally encode the transformed system $A$ and send it back through the channel. This takes time $\frac{2L}{c}$ but requires some large amount of local gates, and has some probability of failure, which means that the implemented quantum operation in the average case is approximately the quantum operation $\EC_U: \rho \ra U \rho U^\dag$.

(ii). If one repeater node can be used, the Protocol 1.1 uses total time $\frac{3L}{2c}$. And this appears to be optimal for exact implementation. If approximate implementation is allowed, we may use Protocol 9, which takes time only $\frac{L}{c}$ but requires some large amount of local gates, and has some probability of failure, which means that the implemented quantum operation in the average case is approximately the quantum operation $\EC_U: \rho \ra U \rho U^\dag$. And Protocol 9 works under the assumption that the error rates in the channels are such that there is positive quantum capacity, unlike the case of other protocols where the allowed error rates can be higher since we use postselection to generate entanglement and then use that to teleport the quantum states.

(iii). If two repeater nodes can be used, the total time cost could be $\frac{7L}{5c}$ under Protocol 1.2($\frac{1}{5},\frac{3}{5}$) or Protocol 3.2($\frac{1}{5},\frac{3}{5}$) (the latter with a suitable choice of the group and representation, as mentioned in Sec.~\ref{sct2}). The required numbers of photons and matter qubits in the protocol are linear in the number of input qubits.  If the $U$ can be written in the fast double-group form or the controlled-group form, with the group being finite, then the total time cost is at most $\frac{4L}{3c}$, under Protocol 5.2($\frac{1}{3},\frac{2}{3}$) in the case of fast double-group form, or Protocol 6.2($\frac{1}{3},\frac{2}{3}$) in the case of controlled-group unitaries. If $U$ can be implemented by any other exact fast protocol, then Protocol 7.2($\frac{1}{3},\frac{2}{3}$) gives a total time cost of $\frac{4L}{3c}$, which is the same as Protocol 5.2($\frac{1}{3},\frac{2}{3}$) or Protocol 6.2($\frac{1}{3},\frac{2}{3}$) above. Note that the entanglement cost is generically linear in the logarithm of the dimension of the Hilbert space that the unitary acts on, which may be different from (and often larger than) the cost for a different unitary acting on the same space which can be written in the fast double-group form or the controlled-group form. This is because the entanglement cost in the double-group or the controlled-group case is linear in the logarithm of the size of the group, which may be much smaller than the dimension of the Hilbert space. For approximate implementation, the Protocol 9 is already optimal in time cost, since $\frac{L}{c}$ is the minimum time to send a message through distance $L$. So any more number of repeater nodes cannot improve the time cost of approximate implementation.

(iv). If three repeater nodes can be used, again since any bipartite unitary can be expanded using the double-group form with a group of size $d_A^2 d_B^2$, the total time cost could be $\frac{7L}{6c}$ under Protocol 1.3($\frac{1}{6},\frac{1}{2},\frac{5}{6}$) or 3.3($\frac{1}{6},\frac{1}{2},\frac{5}{6}$).\\

With more repeater nodes, a smaller time cost could theoretically be achieved, but in practice, the possible errors in the extra local gates and measurements may reduce the benefit of the further small savings in total time cost.

In the following we prove some lower bounds for the total time needed for transmitting information and implementing bipartite unitaries with the help of up to three repeater nodes. The upper bounds stated above will also be combined into the results below. We still use the same assumptions in Sec.~\ref{sct2} which were for implementing unitaries. By ``sending information \emph{unambiguously},'' we mean that the information is sent with probability $p$ where $0<p\le 1$, and in the case the information is not sent (which happens with probability $1-p<1$), the original information is recovered on the original input location, and there is a clear ``flag'' as a measurement outcome in the protocol indicating which of the two cases had happened. The term ``sending classical information'' means implementing a specified classical information channel exactly with capacity strictly greater than zero. This means sending a full bit is not necessary. Similarly, ``sending quantum information'' does not necessarily mean sending a full qubit perfectly, but rather means implementing a specified quantum channel exactly.

\bl\label{le:send}
Suppose $T_s(K)$ is the minimum amount of total time needed to send classical or quantum information unambiguously through a lossy quantum channel from $A$ to $B$ (which are of distance $L$ apart) with the help of $K$ intermediate nodes. Then,\\
(i) $T_s(0)=\frac{2L}{c}.$\\
(ii) $T_s(1)=\frac{4L}{3c}.$\\
(iii) $T_s(2)=\frac{8L}{7c}.$\\
(iv) $T_s(3)=\frac{16L}{15c}.$
\el

The proof is in Appendix~\ref{app:lesend}. Lemma~\ref{le:send} is about the lower bounds for the total time, and the results can be written as the formula
\bea\label{eq:tsn}
T_s(n)=\frac{2^{n+1}}{2^{n+1}-1} \cdot\frac{L}{c},\quad\quad n=0,1,2,3,
\eea
but the proof also contains schemes that achieve such lower bounds. By generalizing the type of schemes in the proof, it is not hard to see that when there are $n$ intermediate nodes, an upper bound for the total time for sending information unambiguously through distance $L$ is given by Eq.~\eqref{eq:tsn}. This quantity is equal to $\frac{L}{c}$ plus the time for light to travel from $A$ to the nearest repeater node located at $\frac{1}{2^{n+1}-1} L$ from $A$ for establishing entanglement between these two nodes. We conjecture that this is the minimum total time for any given integer $n$.\\

Combining Lemma~\ref{le:send} and the protocols analyzed in this paper, we get Theorem~\ref{thm:nodes}. In this theorem, the term \emph{unambiguous implementation} of a bipartite unitary means that the unitary is implemented exactly with probability $p$ where $0<p\le 1$, and in the case the unitary is not implemented (which happens with probability $1-p<1$), the original input state is recovered on the original input parties, and the measurement outcome(s) in the protocol suffice to determine which of the two cases had happened. The background for introducing such definition is as follows: there are approximate protocols of the type of Protocol 9, which may take less total time than the other protocols in this paper which generate entanglement first and then send the data states by teleportation (called \emph{prepare-then-teleport} protocols below). Given Protocol 9, one may wonder why we need to consider ``unambiguous implementation'' at all, since practical applications probably only require approximate implementation. The answer is that Protocol 9 is not suitable for the case that the channels have very high error rates such that the quantum capacity is zero. (Furthermore, sometimes the quantum capacity of channels may even vary in time due to attacks by other parties.) Thus, even if practical applications only require approximate implementation, the prepare-then-teleport protocols would still be needed, and this falls under the category of ``unambiguous implementation.'' In practice, there may be small errors in the prepare-then-teleport protocols, in addition to detectable failure events in entanglement generation which is already allowed. We refer to such protocols as approximate prepare-then-teleport protocols. Such a protocol has the same time cost as the corresponding exact version of the protocol. Thus, although Theorem~\ref{thm:nodes} is stated for the exact prepare-then-teleport protocols, its practical usage is not mainly for this case, but rather for the approximate prepare-then-teleport protocols.

\bt\label{thm:nodes}
Let $T(K)$ be the minimum time that is sufficient for unambiguous implementation of an arbitrarily given bipartite unitary on systems $A$ and $B$ of distance $L$, with the help of $K$ repeater nodes. Then\\
(i) $\frac{2L}{c}\le T(0)\le\frac{3L}{c}.$\\
(ii) $T(1)= \frac{3L}{2c}.$\\
(iii) $\frac{5L}{4c}\le T(2)\le\frac{7L}{5c}.$\\
(iv) $T(3)=\frac{7L}{6c}.$
\et

The proof is in Appendix~\ref{app:thmnodes}.

The results above about the total time for unambiguous implementation of bipartite unitaries are summarized in Table~\ref{tbl_totaltime}, where $n$ is the number of intermediate nodes, and $T_{total}$ is the shortest possible total time. The ``Protocol" column lists the protocol that achieves the upper bound in the previous column. The row labeled $2'$ is for the unitaries that fit the fast double-group form or the controlled-group form with the help of two repeater nodes, see the point (iii) near the beginning of this subsection. Not shown in the table are the bounds of the time cost for a remote single-qubit unitary with the help of up to three repeater nodes with certain restricted parameter choices in the unitary, and the upper bounds coincide with those in the corresponding rows $n=0,1,2,3$. For such tasks with $n>0$, the upper bounds are attained by the variants of Protocol 8 in Sec.~\ref{ssct:3.2}. The upper bound with $n=0$ is attained by Protocol 8 itself. The lower bounds coincide with the $T_s(n)$ in Lemma~\ref{le:send}, since information needs to be sent in at least one direction in any protocol for this task.

\begin{table}
    \begin{tabular}{ | c | c | c |}
    \hline
    $n$ & Range of $T_{total}$ & Protocols attaining the upper bound \\ \hline
    $0$ & $[2\frac{L}{c},3\frac{L}{c}]$ & Protocol 1\\[1ex] \hline
    $1$ & $\frac{3}{2}\frac{L}{c}$ & Protocol 1.1 \\[1ex] \hline
    $2$ & $[\frac{5}{4}\frac{L}{c},\frac{7}{5}\frac{L}{c}]$ & Protocols 1.2($\frac{1}{5},\frac{3}{5}$) and 3.2($\frac{1}{5},\frac{3}{5}$) \\[1ex] \hline
    $2'$ & $[\frac{5}{4}\frac{L}{c},\frac{4}{3}\frac{L}{c}]$ & Protocol m.2($\frac{1}{3},\frac{2}{3}$) with ${\rm m}=5,6,7$ \\[1ex] \hline
    $3$ & $\frac{7}{6}\frac{L}{c}$ & Protocols 1.3($\frac{1}{6},\frac{1}{2},\frac{5}{6}$) and 3.3($\frac{1}{6},\frac{1}{2},\frac{5}{6}$) \\[1ex]   \hline
    \end{tabular}
     \caption{A table showing the possible range of total time cost for unambiguous implementation of bipartite unitaries as a function of the number of repeater nodes $n$. The lower bounds are the best values known but may be improved. The protocols that attain the upper bounds are listed. The row labeled $2'$ is for the fast protocols with two repeater nodes, see text for details.}\label{tbl_totaltime}
\end{table}

The following result shows that the total time cost could approach $\frac{L}{c}$ from above as the number of repeater nodes increases.

\bpp\label{prop:many_nodes}
Let $k$ be a nonnegative integer.\\
(i) There is a protocol for implementing an arbitrary bipartite unitary with the help of $2k+1$ repeater nodes in total time $\left[1+\frac{1}{2(2^{k+1}-1)}\right]\frac{L}{c}$.\\
(ii) There is a protocol for implementing an arbitrary bipartite unitary with the help of $2k$ repeater nodes in total time $\left[1+\frac{2}{2^{k+2}-3}\right]\frac{L}{c}$.\\
\epp
\bpf
(i) The results for the cases $k=0$ and $k=1$ are given by Protocol 1.1 and Protocol 1.3($\frac{1}{6},\frac{1}{2},\frac{5}{6}$), respectively. We may extrapolate Protocol 1.3($\frac{1}{6},\frac{1}{2},\frac{5}{6}$) by adding an even number of repeater nodes in a symmetric configuration. For example, when there are $5$ repeater nodes, they could be at distances $\frac{L}{14},\frac{3L}{14},\frac{L}{2},\frac{11L}{14},\frac{13L}{14}$ from $A$, respectively. The total time cost is $\frac{L}{c}$ plus the time for light to travel from one end node to the nearest repeater node, which is $\frac{L}{14c}$ in this example. For general $k\ge 2$, and number of repeater nodes $n=2k+1$, we let the $(k+1)$-th node be at exactly the middle between the end nodes $A$ and $B$, and choose the distances between neighboring nodes from $A$ to the middle node to be $x L,2x L,4 x L,\dots,2^k x L$, respectively. The remaining nodes are placed at symmetric locations (against the middle node). The total time is the sum of the time for information from $A$ to the middle node which is $(\frac{1}{2}+x)\frac{L}{c}$, plus the time for information to be sent back to $A$ which is $\frac{L}{2c}$. Thus, the total time cost is $(1+x)\frac{L}{c}$. From $2(1+2+\dots+2^k)x=1$, we get $x=\frac{1}{2(2^{k+1}-1)}$, and the total time is $\left[1+\frac{1}{2(2^{k+1}-1)}\right]\frac{L}{c}$. [Note that extrapolating Protocol 3.3($\frac{1}{6},\frac{1}{2},\frac{5}{6}$) gives the same total time cost, while the entanglement cost is generally higher but could be smaller for special classes of $U$.]

(ii) The results for the cases $k=0$ and $k=1$ are given by Protocol 1 and Protocol 1.2($\frac{1}{5},\frac{3}{5}$), respectively. For general $k\ge 2$, and number of repeater nodes $n=2k$, we choose the distances between neighboring nodes from $A$ to the $(k+1)$-th node (counted in the direction from $A$ to $B$) to be $x L,2x L,4 x L,\dots,2^k x L$, respectively, and the distances between neighboring from $B$ to the $(k+1)$-th node to be $y L, 2 y L, \dots, 2^{k-1} y L$, respectively. The time for the information about the input state at $A$ to reach the $(k+1)$-th node is $[1+(1+2+\dots+2^k)] x \frac{L}{c}=2^{k+1}x \frac{L}{c}$. The time for information about the input state at $B$ to reach the $(k+1)$-th node is $[1+(1+2+\dots+2^{k-1})y \frac{L}{c}=2^k y \frac{L}{c}$. We demand that these two quantities should be equal. This implies $y=2x$. Then the spatial configuration of the repeaters implies $(2^{k+1}-1)x+(2^k-1) y =1$. Thus $x=\frac{1}{2^{k+2}-3}$. After the unitary $U$ is done on the $(k+1)$-th node, the parts of the output that belongs to $A$ or $B$ are sent back to the respective end node. This takes time $(1+2+\dots+2^k)x \frac{L}{c}=(2^{k+1}-1)x \frac{L}{c}$.
The total time is $[2^{k+1}+(2^{k+1}-1)]x\frac{L}{c}=(2^{k+2}-1)x\frac{L}{c}=\frac{2^{k+2}-1}{2^{k+2}-3} \frac{L}{c}$. This completes the proof.
\epf

\subsection{Application to position-based quantum cryptography}\label{ssct:3.4}

In the following, we discuss the cases of two and three verifiers (``reference stations'' in the language of \cite{LL11}) respectively. In the former case we shall discuss any number of repeater nodes owned by the attacker, but in the latter case we only discuss a special type of settings.

\textbf{(A). Two verifiers.}

The main observation in the previous literature about the security of quantum position verification is the following: With the fast but approximate unitary protocols for generic bipartite unitaries \cite{Buhrman14,bk11}, quantum position verification could be viewed as insecure (theoretically), when the attacker has only two end nodes and no other nodes. The attacker only has to perform such fast approximate protocol for implementing the bipartite unitary when the states are intercepted by his two end nodes, and after doing a unitary, send the output states to their destination locations. But this does not count in the time for entanglement preparation by the attacker, and another caveat is that the attacker would need to use a large amount of entanglement, generically exponential in the sizes of the input quantum systems for fixed accuracy, which could be unrealistic. So, if such exponential entanglement cost is indeed optimal, quantum position verification could be regarded as secure for practical purposes, as commented in \cite{cl15}.

The lower bounds for the time needed for implementing bipartite unitaries in this paper have immediate implications to position-based quantum cryptography. We consider the one-dimensional case of quantum position verification, which means there are two reference sites (verifiers), and the prover is on the line segment between them (see the corresponding setting in \cite{LL11}). We denote the two reference sites as $V_1$ and $V_2$ and the prover as $P$, where the letter $V$ is for ``verifiers''. The two end nodes of the attacker are still denoted as $A$ and $B$. The implication to be stated below is under the following assumptions (which are to be satisfied simultaneously): \\
(1) The quantum channels between nodes are noisy, but the local operations (including local quantum gates and local measurements) are regarded as having no errors and taking no time to implement, and classical communication is error-free.\\
(2) The speed of light is uniform in the relevant medium and is denoted as $c$.\\
(3) The attacker does not have prior knowledge about the time of the actual run of the position verification scheme, and all nodes of the attacker are notified of the start of the position verification scheme at the same time when the scheme starts, i.e., when quantum states are sent out from one of the verifiers.\\
(4) The quantum memories of all nodes of the attacker have finite decoherence time.\\
(5) The authentic verifiers and prover can establish entanglement regardless of a high loss rate in the channel, by the postselection method in \cite{nemoto14}, and they establish entanglement frequently at all times to guard against decoherence, and so that the time for establishing entanglement need not be counted towards the operating time for (authentic) position verification. The nodes of the attacker cannot do the same, since frequent communication at all times would mean that the activities are easily detected.\\
(6) There is a restricted region of radius $\delta$ centered at the position of the prover in which no nodes of the attacker can reside. (This is also assumed in \cite{LL11}.)\\
(7) The nodes of the attacker, including the end nodes $A$ and $B$, are all located on the line segment $V_1 V_2$, and the prover $P$ is located at the middle point of the line segment $V_1 V_2$.\\

The usual quantum protocols for position verification on a line can be characterized by a unitary $U$ to be carried out on the two systems sent by the two verifiers. (See the formalism in \cite{cl15}, but we ignore the error parameter here.) Note that the assumptions (3) through (5) together imply that the attacker cannot have entanglement prepared at exactly the time needed for the protocol (of course, the attacker could be actively trying to establish entanglement at all times, but then the frequent communication activities would be easy to detect). Under the above assumptions, the usual position verification schemes on a line can be regarded as secure, and the reason is as follows: the total time of the protocol is the sum of the time for sending a quantum state from one verifier $V_1$ to the node $A$ of the attacker, plus the ``total time'' for performing the bipartite unitary on $AB$, plus the time for sending a quantum state from $B$ to $V_2$. The only difference in time compared to the original position verification scheme is caused by the middle time period, i.e. the difference of the time needed for doing the unitary with the time for direct transmission over the distance $L:=\vert AB\vert$. The protocols for generic bipartite unitaries in this paper are the variants of Protocol 1 and Protocol 3. Suppose all used nodes of the attacker are of distance $\delta$ away from the prover $P$. It appears that, under the variants of Protocols 1 or 3, the difference in total time is at least $2\delta/c$. For example, if the Protocol 1.2($\frac{1}{5},\frac{3}{5}$) were used, as the second repeater node is at $\frac{3L}{5}$ from $A$, the $\delta$ is at most $\frac{L}{10}$, so $2\delta/c\le \frac{L}{5}$. But the difference in total time is $\frac{7L}{5c}-\frac{L}{c}=\frac{2L}{5c} > \frac{L}{5c}$. The above examples hint that we should consider the following problem: When the nearest repeater nodes to the prover are at some nonzero distance away from the prover, what is the minimum guaranteed $\Delta t$ (the total time needed to implement a bipartite unitary on $AB$ minus $L/c$)?

For this purpose, it suffices to consider the case that the prover is at exactly the middle of the two verifiers, since, if not, a position verification scheme would need one of the verifiers to send his/her part of the input quantum state to the unitary before the other verifier does so, and effectively the two verifiers are at the same distance to the prover, for the purpose of comparing the time cost of the attacker's scheme of attack and the ideal scheme. If there are no more than three repeater nodes used by the attacker, there is a lower bound: $\Delta t\ge \frac{L}{6c}$, which is from Theorem~\ref{thm:nodes}. If there are more than three repeater nodes, the lower bound of $\Delta t$ should still be nonzero but it would approach zero, as Proposition~\ref{prop:many_nodes} suggests. But in practice the many repeaters would make the attacker's activities easy to detect. The discussion above is summarized by the following postulate.

\begin{postulate}\label{postu:position_verification}
Under the assumptions (1) through (7), quantum position verification with two verifiers and one prover on a line is secure if exact implementation of the unitary is required and there are at most three repeater nodes used by the attacker.
\end{postulate}

The requirement of ``exact implementation'' seems to reduce the significance of the claim, as there are approximate protocols for performing the unitary that use less time when there is a node between $A$ and $B$, such as Protocol 9. We describe Protocol 9 again here for the sake of discussion below: First, use quantum error-correcting codes to encode the input states, and send them through the noisy quantum channels to a middle node, and locally decode and do the unitary there, and encode again and send the states back to the end nodes. In Protocol 9, the middle node is exactly at the middle point of $AB$, but we modify the protocol by relaxing such requirement by moving the middle node a little away from the middle point, for the application to position verification schemes. Such modified Protocol 9 still gives a total time smaller than other protocols. The modified Protocol 9 is only feasible when the error rate of the channels is not too high. In the current case of photonic channels, the error can be modeled as of two types: the loss error and phase error. If the loss-error rate is over some constant, the quantum capacity of the photonic channel would become zero, thus unable to transmit the encoded input state. The photon loss error would not affect the method of entanglement generation in \cite{nemoto14}, since we postselect on the events of photon being detected, thus, the photon is not lost in the transmission. The phase error would not affect the quality of the generated entangled state (see the discussion at the beginning of Sec.~\ref{ssct:3.2}). So the entanglement generation in our main protocols such as the variants of Protocols 1 and 3 are not affected by high error rates in the channel. On the other hand, if we directly send quantum information through such channel, then when the error rate is sufficiently high, the quantum capacity of the channel (over the distance of transmission) becomes zero and thus the modified Protocol 9 becomes inferior to the variants of Protocols 1 and 3. This gives rise to Conjecture~\ref{conj:position_verification} below. The reason why we have a conjecture rather than a definite claim is that there might be some other approximate protocols for performing the unitary, which might work better than both types of protocols above under the same high error rate in the channels.

The term ``approximate quantum position verification'' in Conjecture~\ref{conj:position_verification} below could include two types of approximations, which can appear separately or jointly: one is the approximate implementation of the unitary, and the other is that the position of the prover is not exactly determinable, but could be in some small range. We do allow both types of approximations in the following conjecture. When we say a scheme of approximate quantum position verification is ``secure,'' we mean that the verifiers can statistically distinguish whether the party responding is the attacker(s) or the authentic prover.

\begin{conjecture}\label{conj:position_verification}
Under the assumptions (1) through (7), approximate quantum position verification with two verifiers and one prover on a line is secure, if there are a finite number of repeaters used by the attacker, and the quantum channels between the attacker's repeaters have a sufficiently high error rate over the distance of $2\delta$ so that they have zero quantum capacity.
\end{conjecture}

\begin{figure*}[ht]
\begin{center}
\includegraphics[scale=1]{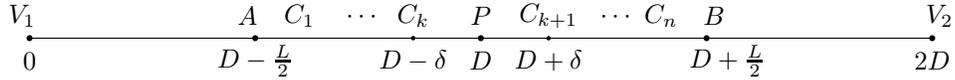}
\end{center}
\caption{An example of spatial configuration for the two-verifier position verification scheme with attacker's nodes $A,B,C_1,\dots,C_n$. Note that $P$ being at the middle point of $AB$ is not required in Proposition~\ref{prop:delta}, but is desirable for the attacker to minimize the total time of acting as the prover. The repeater nodes $C_k$ and $C_{k+1}$ which are neighbors of $P$ are denoted as $E$ and $F$ in the proof of Proposition~\ref{prop:delta}.} \label{fgr_bip}
\end{figure*}

The following result complements the above discussion with a statement about the time difference $\Delta t$. The setup is illustrated in Fig.~\ref{fgr_bip}.

\bpp\label{prop:delta}
Suppose the assumptions (1) through (7) all hold, and furthermore assume the following: the number of repeater nodes of the attacker is arbitrary; the two repeater nodes that are nearest to $P$ are exactly at distance $\delta$ to $P$; the distance $L$ between $A$ and $B$ is larger than $6\delta$. Let $t_{min}$ be the minimum total time needed to implement a bipartite unitary on $AB$, which depends on $L$, $\delta$, the class of allowed protocols, the target set of unitaries, and the required precision. Let $\Delta t=t_{min}-\frac{L}{c}$. Then\\
(i) $\Delta t \ge 2\delta/c$ if exact implementation is required and the target unitary is a generic one, and the protocols are limited to variants of Protocol 1 or Protocol 3;\\
(ii) $\Delta t$ could approach zero as the number of repeater nodes increases, if approximate implementation is acceptable, or if the target unitary is exactly of the fast double-group form or the fast controlled-group-unitary form discussed in Protocols 5 and 6.
\epp

\bpf
(i) The proof is by considering the information flow. To implement a generic bipartite unitary $U$, information has to be sent from $A$ to $B$ and from $B$ to $A$. Denote the two nearest neighbors among the repeater nodes as $E$ and $F$ (denoted as $C_k$ and $C_{k+1}$ in Fig.~\ref{fgr_bip}), where $E$ is nearer to $A$ than $F$ is. By assumption $\vert EF\vert= 2\delta$. If the attacker chooses not to use (one of) $E$ and $F$ in the protocol but uses other repeater node(s) instead, the argument would be similar to what follows but with the new $E$ and $F$ satisfying $\vert EF\vert>2\delta$, giving rise to the same conclusion. Thus we assume the attacker indeed uses $E$ and $F$ in the protocol.  For information from $A$ to be transmitted to $E$ (by teleportation or other means) or backwards, the minimum time needed is $\vert AE\vert/c$, and similarly, for information from $B$ to be transmitted to $F$ or backwards, the minimum time needed is $\vert BF\vert/c$. If $U$ is to be implemented using variants of Protocol 1, some entanglement is to be established between $E$ and $F$, and such time interval could be in the best case contained in the earlier stages of the protocol when states were sent (or teleported) to $E$ and $F$, and after that, some period of time is needed to teleport the state on $E$ to $F$, or from $F$ to $E$, to perform the target unitary on a local party, and teleport the state belonging to the other party back. So in this case the time spent between $EF$ is at least $2\vert EF\vert/c=4\delta/c$. Thus the total time difference $\Delta t$ is at least $2\delta/c$. The case of Protocol 3 is similar, so we abbreviate the argument here.

(ii) By the approximate fast unitary protocols for generic bipartite unitaries \cite{Buhrman14,bk11}, $\Delta t$ could approach zero by the following protocol: First, send or teleport the computational-basis information about the ancilla to $E$ and $F$, and then the implementation of the unitary on $EF$ is  by an approximate fast unitary protocol such as in \cite{Buhrman14} or \cite{bk11} (although this requires a large amount of entanglement for reasonable accuracy in the implementation for generic $U$), which takes time only $\vert EF\vert/c$, and the time interval for preparing entanglement needed for such steps on $EF$ could in the best case be contained in the earlier stages of the protocol when states were sent (or teleported) to $E$ and $F$. When $\vert AE\vert$ and $\vert BF\vert$ are both larger than $2\delta$, and the number of repeater nodes can be increased arbitrarily, the time needed for sending quantum states from $A$ to $E$ and from $B$ to $F$ could approach $\vert AE\vert/c$ and $\vert BF\vert/c$ from above, respectively (see the proof of Proposition~\ref{prop:many_nodes}), and the time for preparing entanglement on $EF$ could be contained in such time period. Thus, when $\vert AB\vert > 6\delta$, and the number of repeater nodes is allowed to vary, the $\Delta t$ could approach zero.

In the case that the target unitary $U$ is exactly of the fast double-group form or the fast controlled-group-unitary form, then the protocol is similar to the above: First, send or teleport the computational-basis information about the ancilla to $E$ and $F$, and perform the core steps of the Protocols 5 and 6 on $E$ and $F$, and send the information about the output systems back through the repeater nodes to $A$ and $B$, and finally perform the local operations in the last steps of the Protocols 5 and 6 to complete the protocol. The $\Delta t$ could approach zero, for the same reason as in the previous paragraph, under the condition that $\vert AB\vert > 6\delta$, and the number of repeater nodes is allowed to vary. This completes the proof.
\epf

We leave open the problem of possible range of $\Delta t$ in the cases not covered in Proposition~\ref{prop:delta}, e.g. the case (i) without restrictions on the protocols.\\

\textbf{(B). Three verifiers.}

It was shown in \cite{LL11} that if there are three verifiers $V_1,V_2,V_3$ and one prover $P$ which are all located in the same plane, and the prover is inside the triangle $V_1 V_2 V_3$, then the pairwise use of the usual two-verifier position verification scheme for each of the three pairs of verifiers would be enough to uniquely locate the prover. In the following we assume the nodes of the attacker are all located in the same plane as the verifiers and the prover, in addition to the assumptions (1) through (6) listed in the two-verifier case. The Figure~\ref{fgr_tripa} below shows an example of spatial configuration of the tripartite position verification scheme, including the attacker's nodes. The $P$ is the location of the prover, and it is at the center of the triangle in this particular example. The locations of the nodes of the attacker depend on which two of the three verifiers are active. In Fig.~\ref{fgr_tripa} the $V_1$ and $V_2$ are active, and thus the end nodes of the attacker, $A$ and $B$, are respectively on the line segment $V_1 P$ and $V_2 P$. The attacker has only one intermediate node in this example, and it is at the point $C$ on the line segment $AB$. We may always assume the two end nodes of the attacker are located on the lines connecting one of the verifiers to the prover, and this is justified by that the quantum state is to be sent from the verifier(s) to the prover in the position verification scheme.

\begin{figure}[ht]
\begin{center}
\includegraphics[scale=1]{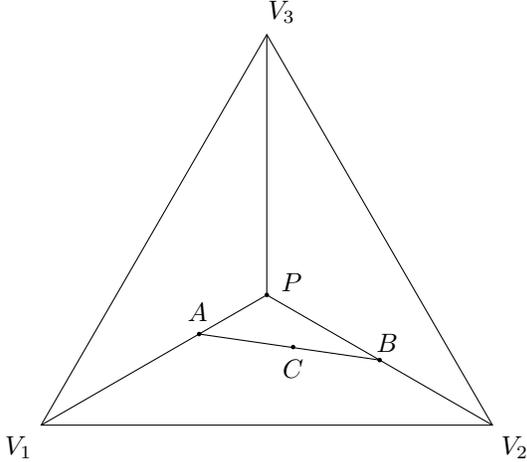}
\end{center}
\caption{An example of spatial configuration for the three-verifier position verification scheme with attacker's nodes $A,B,C$.} \label{fgr_tripa}
\end{figure}

We claim that quantum position verification can be secure under the following conditions: there are three verifiers located at the three vertices of an equilateral triangle, respectively, and the prover is at the center of the equilateral triangle, while the attacker has only three nodes (two end nodes and one intermediate node, and their locations could change for different unitaries with input sent by different verifiers); the assumptions (1) through (6) hold; the exact implementation of the unitaries on input states sent by every pair of verifiers is required. The reason the claim holds is as follows (without loss of generality, assume that the two active verifiers are $V_1$ and $V_2$): The total time for the verifiers $V_1$ and $V_2$ to send signals to the prover and back is $(\vert V_1 P\vert + \vert P V_2 \vert) /c$. On the other hand, the total time for the attacker with two end nodes and one intermediate node is at least $3\vert AB\vert/2c + \vert V_1 A\vert /c + \vert B V_2\vert /c$, where the first term is from the lower bound in Theorem~\ref{thm:nodes}(ii). The total time with the attacker present is always larger than the original scheme without an attacker, since $\frac{3}{2}>\frac{2}{\sqrt{3}}\ge x$, where $x:= (\vert A P\vert + \vert BP\vert)/ \vert AB\vert$. The $x$ attains the maximum value of $\frac{2}{\sqrt{3}}$ when $\vert AP\vert=\vert BP\vert$, but in general could take any value in $[1,\frac{2}{\sqrt{3}}]$.

If the conditions in the previous paragraph are changed so that the triangle is no longer equilateral, or the prover is not at the center of the triangle, we still have the following claim: when the prover is inside the triangle $V_1 V_2 V_3$, and the attacker has at most three nodes including the end nodes, and the three angles $\angle V_1 P V_2$, $\angle V_1 P V_3$, $\angle V_2 P V_3$ are all larger than $2\arcsin\frac{2}{3}$, then the original position verification scheme is secure against such attacker when exact implementation of the unitaries is required.

\smallskip
\subsection{Non-unitary operations}\label{ssct:3.5}

So far we have only considered the implementation of nonlocal unitaries, but in practice non-unitary nonlocal operations may be of interest for position verification protocols and beyond. We mentioned in Sec.~\ref{sct1} that any nonlocal quantum operation can be implemented using nonlocal unitaries followed by local measurements. To see this, note that any nonlocal operation can be modelled by a nonlocal unitary $U$ followed by a nonlocal measurement $M$, and the $M$ could always be modelled by a nonlocal unitary $V$ followed by local measurements of some local subsystems, thus the whole operation is $VU$ followed by local measurements. As a consequence, an upper bound for the time cost of implementing generic bipartite unitaries is also an upper bound for the time cost of implementing bipartite quantum operations.

In general, a bipartite quantum operation on $(d_A\times d_B)$-dimensional space could be modelled as a quantum channel with at most $d_A^2 d_B^2$ Kraus operators, so it may be implemented by performing the whole unitary which models the quantum channel and then ignoring the environment systems. There may be other ways of directly performing nonlocal operations without performing the whole unitary which models the quantum channel. So an interesting direction for further study is to look for protocols that implement some classes of non-unitary nonlocal quantum operations with small costs in time and entanglement, which do not have a directly corresponding protocol for nonlocal unitaries. The following is an example which shows that implementing a bipartite unitary first and then measuring some systems is not always the best way to implement a bipartite non-unitary operation.

In the case of one repeater node, let us consider the problem of implementing a permutation operation with a bipartite quantum input state (in the computational basis) but with classical output. The two parties $A$ and $B$ could measure their input quantum state in the computational basis, and send to the middle party. Some classical operation is done on the middle party, and the outcomes are sent to $A$ and $B$. The total time needed is $\frac{L}{c}$. If a permutation unitary is performed first before measurement, then since some time is spent on entanglement preparation, the total time needed is at least $\frac{3L}{2c}$, according to Theorem~\ref{thm:nodes}(ii).
The corresponding ``unitary'' task is to implement bipartite quantum permutation unitaries (discussed in \cite{cy15,cy16}), and the total time needed including entanglement preparation is also at least $\frac{3L}{2c}$.

\smallskip
\section{Discussions}
\label{sec:discussion}

For implementing a bipartite nonlocal unitary, there are at least two possible ways of using repeaters. The first is to generate entanglement between the end nodes using the repeaters, and then forget about the repeaters and only use LOCC to implement the unitary; the second is to generate entanglement between the neighboring repeater nodes, and such steps may be interspersed or followed by LOCC operations which involve the repeater nodes. Actually, the protocols for implementing bipartite unitaries with at least one repeater node in this paper all use repeaters according to the second way above. The reason that the first way is not discussed explicitly in this paper is that the total time cost is higher than that for the second way.

Now we discuss the local operations done on the repeater nodes, by drawing analogies to those in establishing long-range entanglement. We first briefly review two methods of using repeaters to establish long-range entanglement (for long-distance quantum key distribution or for long-distance quantum communication). The first way, which we call ``entanglement-swapping method,'' is to build entanglement across longer and longer distances based on entanglement swapping, and usually involves many steps of Bell state measurements on the repeater nodes, and classical communication between repeater pairs at different distances (see for example \cite{dlcz01}). The second way, which we call ``relay method,'' is to send an encoded qubit in an entangled pair through the linear array of repeater nodes, with error correction at each node (see for example \cite{msd12}). The second way uses less time than the first way, but the local operations at the repeater nodes are more complex as some error correction operation is performed.

In this paper, the repeater nodes (except one node) are used similarly to the second way (``relay method'') in the previous paragraph. For example, this is the way repeaters are used in Protocol 1.3($\frac{1}{6},\frac{1}{2},\frac{5}{6}$). All repeaters except a middle one send quantum information node-by-node to the middle repeater node, and at that middle node some complex operation corresponding to the target unitary is performed, and the output is sent node-by-node through the array of repeater nodes to the end nodes.

As the example above has shown, our protocols often use complicated operations at one of the repeater nodes, but only limited sets of operations at the other nodes. Such limited operations include Pauli gates and Bell-state measurements. In practice there may be only limited sets of operations or resources available at all repeater nodes, and this may be the case for a network in which the repeaters were originally designed for quantum key distribution. It may be interesting to study which sets of bipartite unitaries can be done under such restrictions about the repeaters, with the same time cost as in the case without such restrictions, but with possibly larger entanglement cost. But the results under such restrictions would be less applicable to position-based quantum cryptography, since it is not very natural to assume that the operations done by the attacker's nodes are limited.

\smallskip
\section{Conclusions}
\label{sct4}

In this paper, we have discussed the total time (including the time for entanglement preparation) for implementing a bipartite unitary with the help of some repeater nodes between the two parties, as well as that for implementing a remote single-qubit unitary with the help of up to three repeater nodes. As mentioned in the Introduction and in Sec.~\ref{ssct:3.2}, such total time is calculated under the idealized scenario that there are enough local ancillary systems for generating entanglement between neighboring nodes successfully. We found lower and upper bounds for the total time as functions of the number of repeater nodes. The upper bound approaches the time for direct one-way signal transmission when the number of repeaters increases. This is because of the way we use the repeaters: most of the time is spent in one-way transmission of information (through piecewise entanglement between neighboring nodes), while a small amount of time is used for preparing entanglement across some segment between two neighboring nodes.

We have ignored the time for doing local gates, since that is about the same for all protocols when local gates are fast compared to the long communication time between the parties. But, all the above is based on the assumption that all nodes are notified of the start of the protocol instantaneously (see the assumptions in Sec.~\ref{sct2}). In general,  if the nodes are not notified of the start of the protocol simultaneously, the total time would be the same or even longer, thus, the lower bounds in Theorem~\ref{thm:nodes} still hold. In the case that the total time is indeed longer, this would be better for the application to position-based quantum cryptography.

We have applied the result on the lower bounds of the total time cost of implementing unitaries to position-based quantum cryptography. Since our lower bounds are greater than $\frac{L}{c}$ when there are at most three nodes owned by the attacker, we claim that the position verification scheme with two verifiers is secure when there are at most three repeater nodes used by the attacker, under some assumptions listed in Sec.~\ref{ssct:3.4}. For some classes of spatial configurations in the three-verifier case, we showed that the pairwise position verification scheme still works under similar assumptions. The other spatial configurations in the three-verifier case need to be studied in more detail. The cases of more than three verifiers also remain to be studied. The problem of implementing non-unitary operations also needs further study, both for the application to position-based quantum cryptography and for other possible applications such as distributed computing where the output is possibly classical. Another open problem is to close the gap between the lower and upper bounds in Theorem~\ref{thm:nodes}.

\smallskip

\section{Acknowledgments} We thank W. J. Munro for providing some references. LY thanks Joseph Fitzsimons and Lin Chen for discussions related to the bipartite Clifford operators. This work has been supported by NICT-A (Japan).

\begin{appendix}

\section{Analysis of Protocol 2.2($x_1,x_2$)}\label{app:ptl22}

Here we consider generic choices of $(x_1,x_2)$ for Protocol 2.2($x_1,x_2$) in order to find the choice with the smallest total time cost.

As the ancillae $a$ and $b$ are located at $C_1$ and $C_2$, respectively, in the following we use $C_1$ and $C_2$ as both names for the location and for the system at such location.

We first consider the case when the two conditions $x_2\le 3x_1$ and $x_1+2x_2\ge 1$ are both satisfied, for reasons to be mentioned below. The controlled-$X^j$ gate on $AC_1$ is to be implemented with the help of entanglement. The first part of the entanglement generation process (sending photons from $C_1$ to $A$) takes time $x_1 \frac{L}{c}$. The second part (to confirm entanglement by sending classical messages from $A$ to $C_1$) coincides with the sending of classical message from $A$ to $C_1$ in the controlled unitary protocol on $A C_1$, which takes time $x_1 \frac{L}{c}$. The first part of entanglement generation between $C_1$ and $C_2$ takes time $(x_2-x_1)\frac{L}{c}$, thus it could be contained within the time periods of the previous steps when $x_2\le 3x_1$. After measurement on $C_1$, the sending of the measurement outcome from $C_1$ to $C_2$, together with the second part of the entanglement generation on $C_1 C_2$, takes time $(x_2-x_1)\frac{L}{c}$. The first part of the entanglement generation on $C_2 B$ takes time $(1-x_2)\frac{L}{c}$, which could be contained in the time interval for above steps when $x_1+2x_2\ge 1$. The second part (to confirm entanglement) could be contained in the time interval for the protocol of implementing the controlled-$V_j$ gate on $C_2 B$ using entanglement on $C_2 B$. Such protocol on $C_2 B$ consumes time $2(1-x_2)\frac{L}{c}$. Then the $C_2$ is measured, and the classical outcome is sent to $A$, which takes time $x_2 \frac{L}{c}$. The total time is $[2x_1 + (x_2-x_1) + 2(1-x_2) + x_2]\frac{L}{c} = (2+x_1)\frac{L}{c}$. The two conditions $x_2\le 3x_1$ and $x_1+2x_2\ge 1$ together imply that $x_1\ge \frac{1}{7}$. Hence the total time is at least $\frac{15L}{7c}$, and the minimum is reached when $x_1=\frac{1}{7}$, $x_2=\frac{3}{7}$.

Next, we consider the case that $x_2\le 3x_1$ holds, but $x_1+2 x_2<1$. The steps up until the message from $C_1$ arrives at $C_2$ take time $(x_1+x_2)\frac{L}{c}$, which is smaller than $(1-x_2)\frac{L}{c}$, thus the steps up until the message from $C_1$ arrives at $C_2$ could be contained in the time interval of the first part of the entanglement generation on $C_2 B$. The latter takes time $(1-x_2)\frac{L}{c}$. The remaining steps take time $[2(1-x_2)+x_2]\frac{L}{c}$ according to the analysis in the previous case. The total time consumption is $(3-2x_2)\frac{L}{c}$, which is greater than $\frac{15L}{7c}$ because $x_2<\frac{3}{7}$, the latter is because if $x_2\ge \frac{3}{7}$, the $x_1$ would be less than $\frac{1}{7}$ since $x_1+2 x_2<1$, then the assumption $x_2\le 3x_1$ does not hold.

Finally, we consider the case that $x_2>3x_1$. This means $x_2-x_1>2x_1$, thus the steps up until the message from $C_1$ arrives at $C_2$ take time $2(x_2-x_1)\frac{L}{c}$. If $2(x_2-x_1)<1-x_2$, then $x_2<\frac{3}{7}$, and the total time would be $[1-x_2 + 2(1-x_2) + x_2]\frac{L}{c}=(3-2x_2)\frac{L}{c}$, which is greater than $\frac{15L}{7c}$. On the other hand, if $2(x_2-x_1)\ge 1-x_2$, the total time would be $[2(x_2-x_1) + 2(1-x_2) + x_2]\frac{L}{c}=(2+x_2-2x_1)\frac{L}{c}$, which is also always not less than $\frac{15L}{7c}$ under the conditions about $x_1,x_2$, since $4 \times x_2>4\times 3x_1$ added to $2(x_2-x_1)\ge 1-x_2$ gives $7x_2-14x_1\ge 1$, hence, $x_2-2x_1\ge \frac{1}{7}$.

Combining the considerations above, the total time is at least $\frac{15L}{7c}$, and the minimum is reached when $x_1=\frac{1}{7}$, $x_2=\frac{3}{7}$.

\section{Analysis of Protocol 3.2($x_1,x_2$).}\label{app:ptl32}

Here we consider generic choices of $(x_1,x_2)$ for Protocol 3.2($x_1,x_2$) in order to find the choice with the smallest total time cost.

We first generate the entanglement between an ancilla $a'$ on $C_1$ and an ancilla on node $A$, and also between an ancilla $b'$ on $C_2$ and an ancilla on node $B$. The first part of the entanglement generation process is sending of photons from node $C_1$ to node $A$, and from node $C_2$ to node $B$. The second part of the entanglement generation process is confirming entanglement, which coincides with the sending of classical messages (from $A$ to $C_1$, and from $B$ to $C_2$) in the first communication step in the protocol for the controlled-$V_A(f)$ [or controlled-$T_B(f)$] gate, which uses the usual protocol for Protocol 4. The above steps on nodes $A C_1$ take time $2 x_1 \frac{L}{c}$. If $x_2-x_1\le 2x_1$, the first part of entanglement generation between $a$ on $C_1$ and $b$ on $C_2$ (which is by sending photons from $C_2$ to $C_1$) is also finished during the above time period, otherwise it partially overlaps with the above operations on $A C_1$ but takes time $(x_2-x_1)\frac{L}{c}$. Then some local measurement is performed on $a$, with the outcome sent classically to node $C_2$ (which is in parallel with sending messages along the same route for confirming entanglement between $a$ and $b$), taking time $(x_2-x_1)\frac{L}{c}$. If the first part of the protocol for the controlled-$T_B(f)$ gate  were finished then, which means $2(1-x_2)\le \max\{2x_1,x_2-x_1\}+(x_2-x_1)$, then it is safe to continue, otherwise they wait until the first part of the protocol for the controlled-$T_B(f)$ gate to finish, which is at time $2(1-x_2)\frac{L}{c}$ from the very beginning. The last part of the protocols for the controlled-$V_A(f)$ [and controlled-$T_B(f)$] gate start to be performed as soon as the corresponding first part finishes, but whether they finish before the end of message transmission from $a$ to $b$ is not important. Then a local correction is done on $b$ according to the received message from $a$, and the gate $\hat C$ is done on $b$, followed by a computational basis measurement of $b$, and the outcome is sent to both end nodes, taking time $\max\{x_2,1-x_2\}\frac{L}{c}$. The protocol is completed by doing local unitary corrections at the end nodes. The total time needed is
\bea\label{eq:time_32}
T&=&[\max\big\{2(1-x_2), \max\{2x_1,x_2-x_1\}+(x_2-x_1) \big\} \notag\\
&& + \max\{x_2,1-x_2\} ] \frac{L}{c}.
\eea
When we take $x_1=\frac{1}{5}$, $x_2=\frac{3}{5}$, we get $T=\frac{7L}{5c}$. If $x_2\le \frac{1}{2}$, we have $T\ge 3(1-x_2)\frac{L}{c}\ge \frac{3L}{2c}>\frac{7L}{5c}$, hence the minimum of $T$ is not achieved when $x_2\le \frac{1}{2}$. So to find the minimum of $T$, we may assume $x_2>\frac{1}{2}$. If $2x_1\ge x_2-x_1$ (i.e. $x_1\ge \frac{x_2}{3}$), we have $T=[\max\{2(1-x_2),x_1+x_2\}+x_2]\frac{L}{c}$. Making use of $x_1\ge \frac{x_2}{3}$, we get $T\ge \max\{2-x_2,\frac{7x_2}{3}\} \frac{L}{c}$, thus the minimum $T$ is $\frac{7L}{5c}$ achieved at $x_2=\frac{3}{5}$ (with $x_1=\frac{1}{5}$) in this case. On the other hand, if  $2x_1< x_2-x_1$ (i.e. $x_1< \frac{x_2}{3}$), we get $T=[\max\{2(1-x_2),2(x_2-x_1)\}+x_2]\frac{L}{c}\ge \max\{2-x_2,\frac{7x_2}{3}\} \frac{L}{c}$, thus the infimum of $T$ is $\frac{7L}{5c}$ but not actually achievable since $x_2=\frac{3}{5}$ and $x_1=\frac{1}{5}$ imply that $x_1=\frac{x_2}{3}$. Combining all cases, we find that the minimum $T$ is $\frac{7L}{5c}$, achieved when $(x_1,x_2)=(\frac{1}{5},\frac{3}{5})$.

\section{Proof of Lemma~\ref{le:send}}\label{app:lesend}

\bpf
The requirement of unambiguous transmission implies that there should not be direct transmission of the input state or the encoded input state. Thus when $K=0$, the transmission is to be carried out through teleportation in the case that the information to be transmitted is quantum; in the case that the information to be transmitted is classical, we can use the one-bit teleportation circuit in \cite{zlc00} for each classical bit to be sent. For $K\ge 1$, the transmission is by stepwise teleportation detailed below.

(i) Consider the case $K=0$. The preparation of entanglement on $AB$ takes two steps: first, send photons from $B$ to $A$ to try to generate entanglement between matter qubits; then, the party $A$ sends a classical signal to party $B$, indicating success and also the information about which atoms were successfully entangled. This second step above could coincide with the teleportation of the input quantum state from $A$ to $B$ (or one-bit teleportation \cite{zlc00} in the case that the task is to transmit classical information), as the latter also involves the sending of classical signals (after some local gates and measurements which are assumed to be fast and accurate). The two time periods of communication above cannot be shortened further, since they are already at light speed. Thus the total time needed is $\frac{2L}{c}$ when $K=0$, and this time is achievable by the protocol above.

(ii) $K=1$. Suppose the intermediate node $E$ is located at distance $x$ from $A$. The steps of a generic scheme are as follows (some steps may start later than stated in the following, but those cases would give worse total time consumption): first, prepare entanglement between $AE$ by sending photons from $E$ to $A$, and then send classical signals from $A$ to $E$, where the latter may coincide with teleportation (or one-bit teleportation in the case that the task is to transmit classical information) of the input state from $A$ to $E$. The above steps take time $2x/c$ in the best case. In the meantime (starting from time zero), entanglement between $EB$ could be prepared by sending photons from $B$ to $E$, and if $x\ge\frac{L}{3}$, this could be finished before the input data state reaches $E$, and the teleportation (or one-bit teleportation in case of classical information) from $E$ to $B$ and also the classical signal for confirming entanglement between $EB$ could start immediately, and the total time needed by the protocol is $(2x+L-x)/c=(L+x)/c$. On the other hand, if $x<\frac{L}{3}$, some wait until time $(L-x)/c$ is needed, and after that the step of sending signals from $E$ to $B$ also takes time $(L-x)/c$, thus the total time of the protocol would be $2(L-x)/c$. Therefore, the minimum is reached when $x=\frac{L}{3}$, and the corresponding total time is $\frac{4L}{3c}$.

The above argument assumes that the information about the input data is at only one spatial location at the end of each step in the protocol. But, this might not hold, since it is conceivable that the information is split into two or more branches at some stage in the protocol, and somehow combined together (by some quantum or classical means) later. If such splitting indeed happens, we may notice that each branch of the information about the input data still needs time at least $\frac{4L}{3c}$ in the current case of one intermediate node, since any branch does contain some non-hidden information, when the value or state in all remaining branches are fixed. Therefore, the whole protocol needs time at least $\frac{4L}{3c}$.

(iii) $K=2$. Suppose the intermediate nodes $E$ and $F$ are located at distance $x$ and $y$ from $A$, respectively, where $0<x<y<L$. The steps of a generic scheme are as follows: first, prepare entanglement between $AE$, by sending photons from $E$ to $A$, and then send classical signals from $A$ to $E$, where the latter coincides with teleportation (or one-bit teleportation in case of classical information) of the input state from $A$ to $E$. These steps take time $2x/c$. In the meantime (starting from time zero), entanglement between $EF$ and between $FB$ could be prepared by sending photons from $F$ to $E$ and from $B$ to $F$, respectively, and if $x\ge\frac{y}{3}$, this could be finished before the input data state reaches $E$, and the teleportation (or one-bit teleportation in case of classical information) from $E$ to $F$ and also the classical signal for confirming entanglement between $EF$ could start immediately, and the time from the start until the input information reaches $F$ is $(2x+y-x)/c=(x+y)/c$. On the other hand, if $x<\frac{y}{3}$, some wait until time $(y-x)/c$ is needed, and after that the step of sending signals from $E$ to $F$ also takes time $(y-x)/c$, thus the time that the input information reaches $F$ would be $2(y-x)/c$ which is larger than $(x+y)/c$ when $x<\frac{y}{3}$. The above means that when $y$ is fixed, the total time until the input information reaches $F$ is a piecewise linear function, with minimum taken at $x=\frac{y}{3}$, and the minimum time is $\frac{4y}{3c}$. By the similar argument, the total time of the entire protocol is a piecewise linear function of $y$ when $x$ is fixed to be $\frac{y}{3}$, and the minimum is reached when $\frac{4y}{3}=L-y$, which means the time until the input information reaches $F$ is the same as the time needed for sending photons from $B$ to $F$ for establishing entanglement on $FB$. Thus, $y=\frac{3L}{7}$ is optimal, and the total time needed by the entire protocol is $(\frac{4y}{3} + L-y)/c$, where the $L-y$ is for sending classical signal for confirming entanglement on $FB$ as well as for sending the classical signal in the (one-bit) teleportation from $F$ to $B$. Therefore, the minimum is reached when $x=\frac{L}{7}, y=\frac{3L}{7}$, and the corresponding total time is $\frac{8L}{7c}$.

Similar to (ii), we do not need to consider the case that the information is split into two or more branches at some stage in the protocol, and somehow combined together later.

(iv) $K=3$. Suppose the intermediate nodes $E$, $F$ and $J$ are located at distance $x, y, z$ from $A$, respectively, where $0<x<y<z<L$. The steps of a generic scheme are as follows: First, prepare entanglement between $AE$, by sending photons from $E$ to $A$, and then send classical signals from $A$ to $E$, where the latter coincides with teleportation (or one-bit teleportation in case of classical information) of the input state from $A$ to $E$. These steps take time $2x/c$. In the meantime (starting from time zero), entanglement on the links $EF$, $FJ$, and $JB$ could be prepared by sending photons from $F$ to $E$, from $J$ to $F$, and from $B$ to $J$, respectively, and if $x\ge\frac{y}{3}$, the sending of photons from $F$ to $E$ could be finished before the input data state reaches $E$, and the teleportation (or one-bit teleportation in case of classical information) and also the classical signal for confirming entanglement between $EF$ could start immediately, and the time from the start until the input information reaches $F$ is $(2x+y-x)/c=(x+y)/c$. On the other hand, if $x<\frac{y}{3}$, by the same argument as in the proof of (iii), the time that the input information reaches $F$ would be $2(y-x)/c$ which is larger than $(x+y)/c$. The above means that when $y$ is fixed, the total time until the input information reaches $F$ is a piecewise linear function of $x$, with minimum taken at $x=\frac{y}{3}$, and the minimum time is $\frac{4y}{3c}$. By the similar argument, when $z$ is fixed, the total time until the input information reaches $J$ is a piecewise linear function of $y$ where $x$ is fixed to be $\frac{y}{3}$, and the minimum is reached when $\frac{4y}{3}=z-y$, which means the time until the input information reaches $F$ is the same as the time needed for sending photons from $J$ to $F$ for establishing entanglement on $FJ$. Thus $y=\frac{3}{7}z$ is optimal, and the minimum time until the input information reaches $J$ is $(\frac{4y}{3} + z-y)/c=\frac{8z}{7c}$, where the $z-y$ is for sending classical signal for confirming entanglement on $FJ$ as well as for sending the classical signal in the (one-bit) teleportation from $F$ to $J$. By the similar argument, given that $L$ is fixed, the total time until the input information reaches $B$ is a piecewise linear function of $z$ where $y$ is fixed to be $\frac{3}{7} z$ and $x$ is fixed to be $\frac{y}{3}=\frac{z}{7}$, and the minimum is reached when $x+z=L-z$, which means the time until the input information reaches $J$ is the same as the time needed for sending photons from $B$ to $J$ for establishing entanglement on $JB$. Thus, $z=\frac{7}{15}L$ is optimal, and the minimum time of the entire protocol is $(\frac{8}{7}z + L-z)/c$, where the $L-z$ is for sending classical signal for confirming entanglement on $JB$ as well as for sending the classical signal in the (one-bit) teleportation from $J$ to $B$. Therefore, the minimum is reached when $x=\frac{L}{15}, y=\frac{L}{5}, z=\frac{7L}{15}$, and the corresponding total time is $\frac{16L}{15c}$.

Similar to (ii) and (iii), we do not need to consider the case that the information is split into two or more branches at some stage in the protocol, and somehow combined together later.

In all cases above, the stated lower bounds of the time costs for sending information are all achievable by the explicit protocols in the proof. This completes the proof.
\epf

\section{Proof of Theorem~\ref{thm:nodes}.}\label{app:thmnodes}

\bpf
We may suppose that the target unitary $U$ is not a product unitary for the purposes of proving bounds for arbitrary $U$, as it takes no time to implement a product unitary, according to our assumptions. Since the size of the input system for $U$ is equal to that of the output system on each party, the non-product unitary $U$ necessarily transmits some (quantum or classical) information about the input state from $A$ to $B$, and some information from $B$ to $A$. To avoid discussing partial qubits, we consider the special case that $U$ is the SWAP gate acting on two qubits in the proof of lower bounds of total time below. The SWAP gate sends one qubit of quantum information in each direction. Since perfect implementation of $U$ is required and there are errors in channels, we cannot transmit information from $A$ to $B$ directly, but have to resort to the use of entanglement, as the possible failure in preparation of entangled states can be remedied by retrying without affecting the data state. Then the data state is sent via teleportation or similar protocols with the help of entanglement, which does not introduce errors since we assume local gates and measurements are error-free. This is the unambiguous way of sending information discussed in Lemma~\ref{le:send}.

(i) Consider the case $K=0$. As mentioned above, some information is to be transferred from $A$ to $B$ for implementing the SWAP gate. Thus a lower bound of time needed for the SWAP gate is given by the quantity $\frac{2L}{c}$ in Lemma~\ref{le:send}(i). We may use Protocol 1 to implement any bipartite unitary $U$ using total time $\frac{3L}{c}$. Thus $\frac{2L}{c}\le T(0)\le\frac{3L}{c}$.

(ii) $K=1$. For proving the lower bound, note that the SWAP gate sends one qubit of quantum information in each direction. Suppose the intermediate node $C$ is placed at distance $xL$ from $A$ on the line interval $AB$. The time for unambiguous sending of quantum information from $A$ to $B$ is $(1+x)\frac{L}{c}$, according to the argument in the proof of Lemma~\ref{le:send}(ii) (the part $x \frac{L}{c}$ is for establishing entanglement between $A$ and $C$). Similarly, the time for unambiguous sending of quantum information from $B$ to $A$ is $[1+(1-x)]\frac{L}{c}$. The maximum of the two quantities above is $(1+\max\{x,1-x\})\frac{L}{c}$, and the minimum of this expression is reached when $x=\frac{1}{2}$, hence, $T(1)\ge\frac{3L}{2c}$.

Putting the intermediate node $C$ at the middle on the line interval $AB$, we may use Protocol 1.1 to implement $U$ using total time $\frac{3L}{2c}$. In the above we have not considered the possibility that the entanglement between some neighboring nodes fails to be created. We call the protocol without such consideration as the \emph{naive} protocol. To fulfill the requirement of unambiguous implementation of $U$, we consider the following enhanced protocol: the entanglement preparation over each link (between neighboring pairs of nodes) is such that either entanglement is prepared with some redundancy (compared to what is required in the naive protocol), or it is regarded as failed. The redundancy is large enough to guarantee that if the entanglement link fails to be established between two nodes, there is enough entanglement in the confirmed links (whose locations are different from the failed link) for the state of the relevant systems (sometimes ancillary systems) to be sent back to their original starting locations via teleportation in order to recover, after some further local gates, the original input state for $U$. In general, the minimum required redundancy is a function of $K$ and the protocol. For the cases of $K\le 3$ and the protocols mentioned in this paper, such redundancy is $1$, meaning that we only need to create one extra ebit for each ebit in the naive protocol. Thus $T(1)\le\frac{3L}{2c}$. Combining with the lower bound above, we have $T(1)=\frac{3L}{2c}$.

(iii) $K=2$. We discuss the upper bound first. The total time cost for implementing an arbitrary bipartite unitary with the help of two repeater nodes is $\frac{7L}{5c}$ under Protocol 1.2($\frac{1}{5},\frac{3}{5}$) or 3.2($\frac{1}{5},\frac{3}{5}$). Similar to (ii), in case some entanglement link fails to be established, there is enough redundant entanglement in the confirmed links for the state of the relevant systems to be sent back via teleportation in order to recover the original input state for $U$. Thus, $T(2)\le \frac{7L}{5c}$.

Now, consider the lower bound of the total time for implementing the SWAP gate unambiguously. Let $C_1, C_2$ be the two intermediate nodes, located at distance $x_1 L$ and $x_2 L$ from $A$ on the line interval $AB$, respectively, where $0<x_1<x_2<1$. Since any lower bound must be not greater than the upper bound $\frac{7L}{5c}$, it must be that the optimal choices of $(x_1,x_2)$ are such that $x_2\ge\frac{3}{5}$, since otherwise the time for unambiguous sending of quantum information from $B$ to $A$ is at least $[1+(1-x_2)]\frac{L}{c}\ge\frac{7L}{5c}$. Thus, in the following we assume $x_2\ge\frac{3}{5}$. By a similar argument, we also assume $x_1\le\frac{2}{5}$.

According to the argument in the proof of Lemma~\ref{le:send}(iii), the time for unambiguous sending of quantum information from $A$ to $C_2$ is $(x_1+x_2)\frac{L}{c}$ when $x_2\le 3x_1$, but is $2(x_2-x_1)\frac{L}{c}$ otherwise. The combined expression is $\max\{x_1+x_2,2(x_2-x_1)\}\frac{L}{c}$. Then, since $x_2\ge\frac{3}{5}$, we have $x_1+x_2\ge 1-x_2$, thus the entanglement generation on $C_2 B$ could be contained in the time interval of the information transfer from $A$ to $C_2$, so the total time for sending information unambiguously from $A$ to $B$ would be $\max\{1+x_1,1+x_2-2x_1\}\frac{L}{c}$.

Similarly, the time for unambiguous sending of quantum information from $B$ to $A$ is $\max\{2-x_2,x_1+2(x_2-x_1)\}\frac{L}{c}$. Since the SWAP gate sends quantum information in both directions, the $T(2)$ is not less than the maximum of the above two quantities. Thus, $T(2)\ge\max\{1+x_1,1+x_2-2x_1,2-x_2,2x_2-x_1\}\frac{L}{c}$, and the minimum of the right-hand side is reached when $x_1=\frac{1}{4}$ and $x_2=\frac{3}{4}$, hence, $T(2)\ge\frac{5L}{4c}$. Combining with the upper bound above, we have $\frac{5L}{4c}\le T(2)\le\frac{7L}{5c}$.

(iv) $K=3$. For the upper bound, the total time cost with the help of three repeater nodes could be $\frac{7L}{6c}$ under Protocol 1.3($\frac{1}{6},\frac{1}{2},\frac{5}{6}$) or 3.3($\frac{1}{6},\frac{1}{2},\frac{5}{6}$). For similar reasons as in (ii) and (iii), in case some entanglement link fails to be established, there is enough redundant entanglement in the confirmed links for the state of the relevant systems to be sent back via teleportation in order to recover the original input state for $U$. Thus, $T(3)\le\frac{7L}{6c}$.

Now, consider the lower bound of the total time for implementing the SWAP gate unambiguously. Let $C_1, C_2, C_3$ be the three intermediate nodes, located at distance $x_1 L$, $x_2 L$, and $x_3 L$ from $A$ on the line interval $AB$, respectively, where $0<x_1<x_2<x_3<1$. Since $T(3)\le\frac{7L}{6c}$, by using an argument similar to that in (iii), we may assume $x_1\le\frac{1}{6}$, and $x_3\ge\frac{5}{6}$.

If $x_2<\frac{1}{2}$, we have $x_3-x_2>2(1-x_3)$, thus the time for sending information unambiguously from $B$ to $A$ is at least $[2(x_3-x_2)+x_2]\frac{L}{c}>\frac{7L}{6c}$, i.e., greater than the known upper bound $\frac{7L}{6c}$ for $T(3)$, thus this case cannot give rise to the optimal lower bound of $T(3)$. Hence, $x_2\ge\frac{1}{2}$. By symmetry, $x_2\le\frac{1}{2}$. Hence, $x_2=\frac{1}{2}$.

Similar to (iii), the time for unambiguous sending of quantum information from $A$ to $C_2$ is $\max\{x_1+x_2,2(x_2-x_1)\}\frac{L}{c}=\max\{x_1+\frac{1}{2},1-2x_1)\}\frac{L}{c}$. Since $x_3\le 3x_2=\frac{3}{2}$, we have that the time for generating entanglement on $C_2 C_3$ could be contained in the time interval for sending information unambiguously from $A$ to $C_2$. Since $x_3\ge\frac{5}{6}$, the time for generating entanglement on $C_3 B$ could be contained in the time interval for sending information unambiguously from $A$ to $C_3$. Hence, the total time for sending information unambiguously from $A$ to $B$ is $\max\{x_1+1,\frac{3}{2}-2x_1\}\frac{L}{c}$. This expression reaches its minimum $\frac{7L}{6c}$ when $x_1=\frac{1}{6}$.

Similarly, the total time for sending information unambiguously from $B$ to $A$ reaches its minimum $\frac{7L}{6c}$ when $x_3=\frac{5}{6}$. Hence $T(3)\ge\frac{7L}{6c}$. Combined with $T(3)\le\frac{7L}{6c}$, we obtain $T(3)=\frac{7L}{6c}$.
\epf

\end{appendix}

\bibliographystyle{unsrt}

\bibliography{channelcontrol}

\end{document}